\documentclass[review]{elsarticle}

\usepackage{lineno}
\modulolinenumbers[5]

\usepackage[utf8]{inputenc} 
\usepackage[T1]{fontenc}    
\usepackage{hyperref}       
\usepackage{url}            
\usepackage{booktabs}       
\usepackage{amsfonts}       
\usepackage{nicefrac}       
\usepackage{microtype}      
\usepackage[utf8]{inputenc} 
\usepackage[T1]{fontenc}    
\usepackage{url}            
\usepackage{booktabs}       
\usepackage{amsfonts}       
\usepackage{nicefrac}       
\usepackage{microtype}      
\usepackage{natbib}
\usepackage{graphicx}
\usepackage{booktabs} 

\usepackage{amssymb,amsthm}
\usepackage{amsmath}
\usepackage{amsfonts}

\usepackage{algorithm}
\usepackage{algpseudocode}

\usepackage{paralist}
\usepackage{multirow}
\usepackage{times}
\usepackage{wrapfig}

\usepackage{url,enumerate}
\usepackage{color,xcolor}
\usepackage{makeidx}  
\usepackage{amsmath,amssymb}
\usepackage[small, compact]{titlesec}
\usepackage{xspace}
\usepackage{epstopdf}
\usepackage{cite}
\usepackage{mathrsfs}
\usepackage{times}
\usepackage{enumerate}
\usepackage{color}
\usepackage{graphicx,epsfig}
\usepackage{grffile}	

\usepackage{amsmath,amssymb,xspace}

\usepackage{paralist}
\usepackage{multirow}
\usepackage{bm}
\usepackage{bbm}
\usepackage{upgreek}
\usepackage{cleveref}
\usepackage{multirow}
\usepackage{subcaption}
\usepackage{ulem}

\usepackage{breqn} 

\usepackage[toc,page]{appendix}

\newcommand{\ours}{{deep coregionalization}\xspace}
\newtheorem{theorem}{Theorem}[section]

\begin{document}

\begin{frontmatter}
\title{Deep Coregionalization for the Emulation of Spatial-Temporal Fields}


\author[sci]{Wei Xing}
\author[sci]{Robert M. Kirby}

\author[cs]{Shandian Zhe\corref{mycorrespondingauthor}}
\cortext[mycorrespondingauthor]{Corresponding author}
\ead{zhe@cs.utah.edu}

\address[sci]{Scientific Computing and Imaging Institute, University of Utah}
\address[cs]{School of Computing, University of Utah}


\begin{abstract}  
Data-driven surrogate models are widely used for applications such as design optimization and uncertainty quantification, where repeated evaluations of an expensive simulator are required.
For most partial differential equation (PDE) simulators, the outputs of interest are often spatial or spatial-temporal fields, leading to very high-dimensional outputs. 
Despite the success of existing data-driven surrogates for high-dimensional outputs, most methods require a significant number of samples to cover the response surface in order to achieve a reasonable degree of accuracy. 
This demand makes the idea of surrogate models less attractive considering the high computational cost to generate the data.
To address this issue, we exploit the multi-fidelity nature of a PDE simulator and introduce \textit{deep coregionalization}, a Bayesian non-parametric autoregressive framework for efficient emulation of spatial-temporal fields.
To effectively extract the output correlations in the context of multi-fidelity data, we develop a novel dimension reduction technique, residual principal component analysis.
Our model can simultaneously capture the rich output correlations and the fidelity correlations and make high-fidelity predictions with only a few expensive, high-fidelity simulation samples.
We show the advantages of our model in three canonical PDE models and a fluid dynamics problem. The results show that the proposed method cannot only approximate a simulator with significantly less cost (at bout 10\%-25\%) but also further improve model accuracy. 

%

\end{abstract}


\end{frontmatter}

\newcommand{\var}{{\rm Var}}
\newcommand{\Tr}{^{\rm T}}
\newcommand{\vtrans}[2]{{#1}^{(#2)}}
\newcommand{\kron}{\otimes}
\newcommand{\schur}[2]{({#1} | {#2})}
\newcommand{\schurdet}[2]{\left| ({#1} | {#2}) \right|}
\newcommand{\had}{\circ}
\newcommand{\diag}{{\rm diag}}
\newcommand{\invdiag}{\diag^{-1}}
\newcommand{\rank}{{\rm rank}}
\newcommand{\nullsp}{{\rm null}}
\newcommand{\tr}{{\rm tr}}
\renewcommand{\vec}{{\rm vec}}
\newcommand{\vech}{{\rm vech}}
\renewcommand{\det}[1]{\left| #1 \right|}
\newcommand{\pdet}[1]{\left| #1 \right|_{+}}
\newcommand{\pinv}[1]{#1^{+}}
\newcommand{\erf}{{\rm erf}}
\newcommand{\hypergeom}[2]{{}_{#1}F_{#2}}

\renewcommand{\a}{{\bf a}}
\renewcommand{\b}{{\bf b}}
\renewcommand{\c}{{\bf c}}
\renewcommand{\d}{{\rm d}}  
\newcommand{\e}{{\bf e}}
\newcommand{\f}{{\bf f}}
\newcommand{\g}{{\bf g}}
\newcommand{\h}{{\bf h}}
\renewcommand{\k}{{\bf k}}
\newcommand{\m}{{\bf m}}
\newcommand{\mb}{{\bf m}}
\newcommand{\n}{{\bf n}}
\renewcommand{\o}{{\bf o}}
\newcommand{\p}{{\bf p}}
\newcommand{\q}{{\bf q}}
\renewcommand{\r}{{\bf r}}
\newcommand{\s}{{\bf s}}
\renewcommand{\t}{{\bf t}}
\renewcommand{\u}{{\bf u}}
\renewcommand{\v}{{\bf v}}
\newcommand{\w}{{\bf w}}
\newcommand{\x}{{\bf x}}
\newcommand{\y}{{\bf y}}
\newcommand{\z}{{\bf z}}
\newcommand{\A}{{\bf A}}
\newcommand{\B}{{\bf B}}
\newcommand{\C}{{\bf C}}
\newcommand{\D}{{\bf D}}
\newcommand{\E}{{\bf E}}
\newcommand{\F}{{\bf F}}
\newcommand{\G}{{\bf G}}
\renewcommand{\H}{{\bf H}}
\newcommand{\I}{{\bf I}}
\newcommand{\J}{{\bf J}}
\newcommand{\K}{{\bf K}}
\renewcommand{\L}{{\bf L}}
\newcommand{\M}{{\bf M}}
\newcommand{\N}{\mathcal{N}}  
\newcommand{\MN}{\mathcal{MN}} 
\newcommand{\Acal}{\mathcal{A}}
\newcommand{\Ocal}{\mathcal{O}}
\newcommand{\Dcal}{\mathcal{D}}
\newcommand{\Ycal}{\mathcal{Y}}
\newcommand{\Zcal}{\mathcal{Z}}
\newcommand{\Fcal}{\mathcal{F}}
\newcommand{\Vcal}{\mathcal{V}}
\newcommand{\Lcal}{\mathcal{L}}
\newcommand{\Tcal}{\mathcal{T}}
\newcommand{\Gcal}{\mathcal{G}}
\newcommand{\Hcal}{\mathcal{H}}
\newcommand{\Scal}{\mathcal{S}}

\renewcommand{\O}{{\bf O}}
\renewcommand{\P}{{\bf P}}
\newcommand{\Q}{{\bf Q}}
\newcommand{\R}{{\bf R}}
\renewcommand{\S}{{\bf S}}
\newcommand{\T}{{\bf T}}
\newcommand{\U}{{\bf U}}
\newcommand{\V}{{\bf V}}
\newcommand{\W}{{\bf W}}
\newcommand{\X}{{\bf X}}
\newcommand{\Y}{{\bf Y}}
\newcommand{\Z}{{\bf Z}}
\newcommand{\Mcal}{{\mathcal{M}}}
\newcommand{\Wcal}{{\mathcal{W}}}
\newcommand{\Ucal}{{\mathcal{U}}}

\newcommand{\bfLambda}{\boldsymbol{\Lambda}}

\newcommand{\bsigma}{\boldsymbol{\sigma}}
\newcommand{\balpha}{\boldsymbol{\alpha}}
\newcommand{\bpsi}{\boldsymbol{\psi}}
\newcommand{\bphi}{\boldsymbol{\phi}}
\newcommand{\boldeta}{\boldsymbol{\eta}}
\newcommand{\Beta}{\boldsymbol{\eta}}
\newcommand{\btau}{\boldsymbol{\tau}}
\newcommand{\bvarphi}{\boldsymbol{\varphi}}
\newcommand{\bzeta}{\boldsymbol{\zeta}}
\newcommand{\bepsilon}{\boldsymbol{\epsilon}}

\newcommand{\blambda}{\boldsymbol{\lambda}}
\newcommand{\bLambda}{\mathbf{\Lambda}}
\newcommand{\bOmega}{\mathbf{\Omega}}
\newcommand{\bomega}{\mathbf{\omega}}
\newcommand{\bPi}{\mathbf{\Pi}}

\newcommand{\btheta}{\boldsymbol{\theta}}
\newcommand{\bpi}{\boldsymbol{\pi}}
\newcommand{\bxi}{\boldsymbol{\xi}}
\newcommand{\bSigma}{\boldsymbol{\Sigma}}

\newcommand{\bgamma}{\boldsymbol{\gamma}}
\newcommand{\bGamma}{\mathbf{\Gamma}}

\newcommand{\bmu}{\boldsymbol{\mu}}
\newcommand{\1}{{\bf 1}}
\newcommand{\0}{{\bf 0}}

\newcommand{\bs}{\backslash}
\newcommand{\ben}{\begin{enumerate}}
\newcommand{\een}{\end{enumerate}}

 \newcommand{\notS}{{\backslash S}}
 \newcommand{\nots}{{\backslash s}}
 \newcommand{\noti}{{\backslash i}}
 \newcommand{\notj}{{\backslash j}}
 \newcommand{\nott}{\backslash t}
 \newcommand{\notone}{{\backslash 1}}
 \newcommand{\nottp}{\backslash t+1}

\newcommand{\notk}{{^{\backslash k}}}
\newcommand{\notij}{{^{\backslash i,j}}}
\newcommand{\notg}{{^{\backslash g}}}
\newcommand{\wnoti}{{_{\w}^{\backslash i}}}
\newcommand{\wnotg}{{_{\w}^{\backslash g}}}
\newcommand{\vnotij}{{_{\v}^{\backslash i,j}}}
\newcommand{\vnotg}{{_{\v}^{\backslash g}}}
\newcommand{\half}{\frac{1}{2}}
\newcommand{\msgb}{m_{t \leftarrow t+1}}
\newcommand{\msgf}{m_{t \rightarrow t+1}}
\newcommand{\msgfp}{m_{t-1 \rightarrow t}}

\newcommand{\proj}[1]{{\rm proj}\negmedspace\left[#1\right]}
\newcommand{\argmin}{\operatornamewithlimits{argmin}}
\newcommand{\argmax}{\operatornamewithlimits{argmax}}

\newcommand{\dif}{\mathrm{d}}
\newcommand{\abs}[1]{\lvert#1\rvert}
\newcommand{\norm}[1]{\lVert#1\rVert}

\newcommand{\ie}{{i.e.,}\xspace}
\newcommand{\eg}{{e.g.,}\xspace}
\newcommand{\etc}{{etc.}\xspace}

\newcommand{\EE}{\mathbb{E}}
\newcommand{\dr}[1]{\nabla #1}
\newcommand{\VV}{\mathbb{V}}
\newcommand{\sbr}[1]{\left[#1\right]}
\newcommand{\rbr}[1]{\left(#1\right)}
\newcommand{\cmt}[1]{}

\newcommand{\bi}{{\bf i}}
\newcommand{\bj}{{\bf j}}
\newcommand{\bK}{{\bf K}}
\newcommand{\Vtr}{\mathrm{Vec}}

\newcommand{\cov}{{\rm Cov}}	

\newtheorem{theore}{Theorem}
\newtheorem{propos}{Proposition}
\newtheorem{lemma}{Lemma}
\newtheorem{coroll}{Corollary}
\newtheorem{remark}{Remark}

\newcommand{\RR}{\mathbb{R}}

\section{Introduction}
Applications such as uncertainty quantification, design optimization, and
inverse parameter estimation demand repeated simulations of partial differential equations (PDEs) in an input parameter space~\citep{Bilionis,Keane}.
Due to the high computational cost of a simulation, a data-driven surrogate model, also known as an emulator, is often employed in place of the simulator~\citep{Kennedy2000,Santner2003}.
In practice, simulations of PDEs generally produce large spatial or spatial-temporal fields (e.g., velocity, temperature, or electric fields).
Due to the large size, modeling the field results directly poses a huge challenge in terms of model capacity and scalability for the surrogate model~\citep{higdon2008computer,xing2015reduced,xing2016manifold}.

Gaussian process (GP) modeling is one of the most commonly used data-driven surrogates due to its capability to (1) quantify model uncertainty, (2) adopt prior knowledge, and (3) avoid overfitting for small datasets~\citep{Kennedy2000,santner2003design,Kennedy2006,Rougier2009,Tagade2013,Lee2013}. Unfortunately, GPs do not naturally extend to multiple output scenarios, especially the high-dimensional simulator spatial-temporal fields.
Modeling the high-dimensional outputs entails efficiently learning the output correlations.
A naive approach is to simply treat the output indexes (indicating the output value at a particular location of the spatial-temporal domain) as additional input parameters~\citep{kennedy2001bayesian}.
This approach is infeasible for high-dimensional outputs since the computational complexity for a general GP is $\mathcal{O}(N^3d^3)$, where $N$ is the number of training samples and $d$ is the output dimensionality~\citep{McFarland2008}.
If we assume a separable structure between the input correlations and the output correlations, the high computational issues could be efficiently circumvented.
Such a relaxation gives rise to the linear model of coregionalization (LMC), the classical framework for multi-output regression~\citep{matheron1982pour,goulard1992linear}.
LMC linearly combines base functions with latent processes to model the high-dimensional output.
Many modern multi-output GP models can be considered special instances or variants of LMC~\citep{Konomi2014,Fricker2013,Rougier2008,zhe2019scalable}.
\citet{Conti2010} proposed a simplified LMC model, namely, the intrinsic coregionalization model (ICM)~\citep{Wackernagel}, which assumes a single latent process with a correlation matrix to govern the output correlations.
To enhance the flexibility of ICM, \citet{higdon2008computer} found a set of bases from singular value decomposition (SVD) on the training outputs to encode the complicated spatial-temporal correlations for PDE simulation problems. This approach is further extended by using nonlinear dimension reduction techniques, such as Kernel PCA~\citep{xing2016manifold} and Isomap~\citep{xing2015reduced} for nonlinear correlations.
Recent developments of GP further introduce latent coordinate features and tensor algebra to improve the model flexibility and scalability~\citep{zhe2019scalable,wilson2015kernel}.
Other methods that relax the separable assumption include \textit{process convolution}~\citep{higdon2002space, boyle2005dependent,teh2005semiparametric, bonilla2008multi} and deep learning hybrid GPs~\citep{wilson2016deep,wilson2011gaussian}. These methods, however,  either are non-scalable to very high-dimensional outputs or require massive tuning. They are thus not suitable for emulations of spatial-temporal simulator fields.

Despite the success of LMC-based methods for emulations of spatial-temporal fields, in order to achieve a reasonable degree of accuracy, they require a large number of samples (corresponding to different inputs) that adequately cover the response surface.
We will further show that the LMC model expressiveness is indeed limited by the number of training samples.
As the number of required training samples grows drastically for a highly non-linear response surface, the idea of data-driven surrogate modeling becomes less practical.

From the data-driven surrogate perspective, an efficient way to relax the high demand of training data is to inject prior knowledge of the PDE models into surrogate models, \eg physical informed neural networks~\citep{raissi2019physics}, conservation kernels~\citep{macedo2010learning}, and hybrid physics-based data-driven surrogates~\citep{shah2017reduced}.
Despite their success, these approaches cannot generalize effectively to the emulation of spatial-temporal fields. Neural networks based method require massive model tunning and are prone to overfitting; conservation kernels only apply to specific types of PDEs; hybrid physics methods require modification of original simulation codes as an intrusive method.
Another more common solution for the emulation is to take the advantage of the multi-fidelity nature of a simulator~\citep{peherstorfer2018survey,kennedy2000predicting,perdikaris2017nonlinear,cutajar2019deep} and fuse information from different fidelities.
More specifically, when generating training data for the data-driven surrogate models, we can easily reduce the computational cost by using a low-fidelity simulator (e.g., a simulator with a sparse discretization mesh) at the price of getting less accurate samples.
The low-fidelity samples, despite being noisy and biased, normally show a strong correlation with the high-fidelity samples.
Thus, the low-fidelity samples can be used in numerous ways to accelerate many science and engineering processes~\citep{peherstorfer2018survey}
In this paper, we focus on how to harness the fidelity correlations such that we can approximate the simulator with many affordable low-fidelity samples and transfer the knowledge to predict highly accurate results without full reliance on expensive high-fidelity data.
To this end, we propose deep coregionalization, an efficient data-driven surrogate for emulations of very high-dimensional spatial-temporal fields.
Our model simultaneously captures the correlation in the spatial-temporal fields and the correlations between different fidelity observations to provide accurate high-fidelity predictions.
Our contributions are as follows:
\begin{itemize}
	\item 
	We prove that under a mild assumption of the kernel structure, the propagation of fidelity knowledge of spatial-temporal fields could be done efficiently though a univariate GP autoregressive model of latent processes.
	\item
	To improve the model capacity for complex problems, we introduce deep coregionalization, which generalizes the GP auto-regressive model for spatial-temporal fields by integrating into the classic linear model of coregionalization (LMC). 
	\item 
	To efficiently capture the output correlations of different fidelities without introducing an over-complicated model with excessive parameters, 
	we propose residual principal component analysis (ResPCA), a dimension reduction technique that can effectively encode the rich correlations of high-dimensional fields, efficiently preserve a compact representation for multi-fidelity fields, and easily scale to high-dimensional field outputs (\eg 1 million).
	\item
	We validate the proposed method on three canonical PDEs (Poisson's equation, Burger's equation, and the heat equation) and a fluid dynamics problem. The results show a significant improvement not only on reducing the simulation budget (for generating the training data) but also on improving the predictive accuracy for the high-fidelity results, compared with models based on single fidelity data.
	
\end{itemize}
The rest of this paper is organized as follows:
Section 2 makes clear the problem of the emulation of spatial-temporal fields .
Section 3 discusses the background, including univariate GP, LMC, and GP autoregression for emulation problems. It then combines the LMC and the GP autoregression to derive the proposed deep coregionalization model.
After some discussions on the model training, ResPCA algorithm is introduced to extract compact representations of the multi-fidelity data.
Section 4 demonstrates the superiority of the proposed method compared with the state-of-the-art high-dimensional GP emulators on three canonical PDEs---Poisson's equation, Burger's equation, and the heat equation---and a fluid dynamics problem, the lid-driven cavity. 
Section 5 concludes this paper with discussions on deep coregionalization's connections to the existing models and further improvements.


\cmt{
	The seminal work of \citep{kennedy2000predicting} proposes an autoregressive method that assumes the function of each fidelity to be a linear transform from the previous fidelity.
	This method is further extended by replacing the linear transformation with a GP regression to model more complicate problems~\citet{perdikaris2017nonlinear}.
	To fully capture and propagate the uncertainty throughout all fidelities, \citet{cutajar2019deep} jointly learned this chain GP model in the deep GP framework~\citep{damianou2013deep}.
	Despite of their great success, existing multi-fidelity data-driven surrogates apply to only simple problems where the output is a scalar value. 
	In practice, simulations of PDEs generally produce large spatial or spatial-temporal field results (e.g., velocity, temperature, or electric fields); the quantity of interest is often the entire spatial-temporal field\citep{higdon2008computer,xing2015reduced,xing2016manifold}.
	In this case, the existing multi-fidelity methods cannot directly apply to high-dimensional field problem because the input for the consequent fidelity GP would have excessively dimension (\eg 1 million).
	Besides, it is unclear how to model the rich correlation between the high-dimensional field outputs to improve the model performance.
}
\section{Statement of the Problem}\label{emulate}
Consider a parameterized nonlinear system of steady-state or transient PDEs of arbitrary order for dependent variables (scalar fields) $u(\x,\s ,t)$, where $\x\in \mathbb{R}^l$ is a vector that parameterizes the PDEs, $\s$ is the spatial index, and $t$ denote the temporal index.
The PDEs can be fully nonlinear and parameterized in an arbitrary fashion (including the initial and boundary conditions); they are also assumed to be well-posed (\ie solutions always exist and are unique) for the range of values of $\x$ considered.

The quantity of interest is the function $u(\x,\s,t)$ in the given range of $\x$.
Direct approximation of this function is difficult due to the amount of samples we need to cover the response surface for such a high-dimensional space~\citep{higdon2008computer}.
Instead, we can record values at specified (fixed) spatial locations, $\s_i$, $i=1,\dots,d_s$, and temporal locations, $t_i$, $i=1,\dots,d_t$.
For different input parameters $\x_i\in \mathbb{R}^l$, the  outputs of the simulator are represented as vectors: $\y^{(F)}_i = \y^{(F)}(\x_i) = (u(\x_i,\s_1,t_1),...,u(\x_i,\s_{d_s},t_{d_t}))^T \in \mathbb{R}^d$, where $d=d_s\times d_t$. 
Here, we use the superscript ${(F)}$ to denote that the resulting vector is from an accurate, high-fidelity simulation.
The value at different spatial-temporal locations can be achieved easily through interpolation provided that the spatial-temporal fixed points are dense enough; interpolation in this case comes with ignorable extra computational effort~\citep{higdon2008computer}.
With the preceding process, the simulation is now treated as a mapping from $\mathbb{R}^l$ to $\mathbb{R}^d$, where $d$ is extremely large.
The goal of a data-driven surrogate is to approximate the mapping given training set $\mathcal{D} = \{\X, \Y^{(F)}\}$, where $\X = [\x_1,\ldots,\x_N]^T$ and $\Y^{(F)} = [\y^{(F)}_1,\ldots,\y^{(F)}_N]^T$.
Provided that we have adequate number of samples to cover the response surface, we can approximate the mapping accurately.

As discussed in the introduction, we intend to avoid relying on only high-fidelity data. 
The low-fidelity data can be cheaply obtained by reducing the fidelity setting (\eg the number of nodes of a discretization mesh and the order of basis functions of the finite element method) of a simulator.
In general, we can run a simulation at different fidelity settings to obtain a multi-fidelity dataset $\{\X^{(f)}, \Y^{(f)}\}_{f=1}^F$, where $\Y^{(f)}$ is a $N_f\times d$ matrix for observations at fidelity-$f$ with $\X^{(f)}$ being an $N_f\times l$ matrix for the corresponding inputs.
We use $f=F$ to denote the highest fidelity and $f=1$ the lowest.
Following the common setting for multi-fidelity data in~\citep{perdikaris2017nonlinear,cutajar2019deep}, we also require that the training inputs of fidelity $f$ is a subset of the previous fidelity $f-1$, .i.e, $\X^{(f)} \subset \X^{(f-1)}$. For convenience, 
we introduce the index notation $\e$ to extract rows form a standard matrix such that $\X^{(f)} = \X_{\e_f,:}$, where $\e_f$ indicates the subset indexes for fidelity data $f$ and $\X$ contains all candidates.
Our goal of surrogate model becomes: given multi-fidelity data $\{\X^{(f)}, \Y^{(f)}\}_{f=1}^F$, how do we efficiently approximate the mapping between $\x$ and $\y^{(F)}$.

%


\section{Model Formulation}

\subsection{Univariate Gaussian Process}
\label{sec:sgp}
We first review the GP regression for learning a univariate function using a high-fidelity dataset $\Dcal = \{\X, \Y^{(F)}\}$.
The quantity of interest is assumed univariate, i.e., $d=1$. We use $y^{(F)}_i$ to distinguish a univariate observation from a multivariate one $\y^{(F)}_i$.
A GP assumes that the finite set of function values on $\X = [\x_1, \ldots, \x_N]^\top$, namely $\y^{(F)} = [y^{(F)}_1, \ldots, y^{(F)}_N]^\top$, follows a multivariate Gaussian distribution, 
$p(\y^{(F)} | \X) = \N(\y^{(F)} | \m, \K + \tau^{-1} \I)$.
Here $\m = [m(\x_1), \ldots, m(\x_n)]^\top$ is the mean function that is usually set to $\0$ after centering $\Y^{(F)}$, and 
$\tau$ is the inverse noise variance, which accounts for model inadequacies and numerical error~\citep{kennedy2001bayesian,rasmussen2003gaussian}.
$\K$ is the covariance matrix with elements $[\K]_{ij} = k(\x_i, \x_j)$, where $k(\x_i, \x_j)$ is a kernel function for the corresponding inputs.
Choosing the right kernel function for a specific application is nontrivial. 
When there is no prior knowledge to guide the choice, the automatic relevance deterrence (ARD) kernel \citep{rasmussen2003gaussian},
\begin{equation}
	k(\x_i, \x_j) = \theta_0 \exp\left(-(\x_i-\x_j) \diag(\theta_1,\ldots,\theta_l) (\x_i-\x_j)^T \right),
\end{equation}
is often utilized. The ARD kernel can freely capture the individual influence of each input parameter on the output results.
The hyperparameters $\{\tau,\theta_0, \ldots, \theta_l\}$ can be estimated by maximizing  the  marginal likelihood of the GP regression,
\begin{equation}
\label{eq:sgp likelihood}
\Lcal = \frac{1}{2} \ln|\K + \tau^{-1} \I| - \frac{1}{2}{\y^{(F)}}^T (\K+\tau^{-1} \I)^{-1} \y^{(F)} - \frac{N}{2} \ln(2\pi).
\end{equation}
The main computational cost is the inversion of $\K$, which is $\Ocal(N^3)$ and $\Ocal(N^2)$ for time and space complexity, respectively.
Given a new input $\x_*$, we can derive its posterior using a conditional Gaussian distribution,
\begin{equation}
\label{eq:sgp}
\begin{aligned}
 p\big(y^{(F)}_*|\x_*, \X,\y\big) &= \N\left(y^{(F)}_* | \mu(\x_*), v(\x_*)\right) \\
 \mu(\x_*) &= \k_{*}^\top (\K + \tau^{-1} \I)^{-1}\y^{(F)} \\
 v(\x_*) &= k(\x_*, \x_*) - \k_{*}^\top(\K + \tau^{-1}\I)^{-1}\k_{*},
\end{aligned}
\end{equation}
where, $\k_{*} = [k(\x_*, \x_1), \ldots, k(\x_*, \x_{N_F})]^\top$ is the vector of covariance between $\x_*$ and $\X$. 


\subsection{Linear Model of Coregionalization}
In order to model the high-dimensional output correlations within limited computational resources, the linear model of coregionalization~\citep{matheron1982pour} assumes a separable decomposition structure,
\begin{equation} \label{eq:lmc}
\y^{(F)}(\x) = \sum_{r=1}^{R_F} \b^{(F)}_r z^{(F)}_r(\x) = \B^{(F)} \z^{(F)}(\x). 
\end{equation}
In this formulation, 
$\B^{(F)} = [\b^{(F)}_1,\dots,\b^{(F)}_{R_f}]$ is the collection of bases $\b^{(F)}_r \in \mathbb{R}^d$; $\z^{(F)}(\x) = [\z^{(F)}_1(\x),\dots,\z^{(F)}_{R_F}(\x)]^T$ is the collection of corresponding latent processes, \ie GPs.
It is assumed that the latent processes are independent (i.e., $\cov( z^{(F)}_r(\x), z^{(F)}_{r'}(\x')) = 0$ for $r \neq r'$) and the bases are orthogonal (${\B^{(F)}}^T \B^{(F)}=\I$).
The LMC model essentially assumes several independent latent processes to fully characterize the data variability. 
The model has been shown particularly effective for the emulations of fields problems~\citep{higdon2008computer}.
The formulation implicitly places the following GP prior:
\begin{equation}
\label{eq:lmc in sum}
\y^{(F)}(\x) = \mathcal{N}(\0, \sum_{r=1}^{R_F} b_r b_r^T k_r(\x, \x') ).
\end{equation}
Given the high-fidelity observations $\{\X, \Y^{(F)}\}$, we can derive the posterior
\begin{equation}
\label{eq:lmc post}
\begin{aligned}
p\big(\y^{(F)}_*|\x_*, \X,\y\big) &= \N\left(\y^{(F)}_* | \bmu(\x_*), \v(\x_*)\right) \\
\bmu(\x_*) &= \sum_{r=1}^{R_F} \b_r^{(f)} \k_{r*}^\top (\K_r + \tau_r^{-1} \I)^{-1}\Y^{(F)}\b_r^{(f)} \\
\v(\x_*) &= \sum_{r=1}^{R_F} \b_r^{(f)} \left(k_r(\x_*, \x_*) - \k_{r*}^\top(\K_r + \tau_r^{-1}\I)^{-1}\k_{r*}\right) {\b_r^{(f)}}^T,
\end{aligned}
\end{equation}
where $\k_{r*} = [k_r(\x_*, \x_1), \ldots, k_r(\x_*, \x_{N_F})]^\top$ is the vector of covariance between $\x_*$ and $\X$. 
We can see that flexibility of the posterior of $\y^{(F)}_*$ is heavily determined by the size of $\K_r$, which is the number of high-fidelity samples $N_f$. 

\subsection{Deep Gaussian Process Autoregression}
Before we move to our model, we make a brief review of deep GP auto-regression, which is used later to derive our model.
Let us consider the multi-fidelity, univariate data,
\ie $\{\X^{(f)},\Y^{(f)}\}_{f=1}^F$ where $\Y^{(f)} = [y^{(f)}_1,\ldots,y^{(f)}_{N_f}]^T $ is a collection of samples at fidelity $f$.
The general auto-regressive formulation for multi-fidelity data is
\begin{equation}
y^{(f)}(\x) = g^{(f)}\left(y^{(f-1)}(\x)\right) + \epsilon^{(f)},
\end{equation}
where $g^{(f)}(y^{(f-1)}(\x))$ is an arbitrary function that maps the low-fidelity results to the high-fidelity ones and $\epsilon^{(f)}$ captures the model inadequacies and numerical errors.
If we assume a simple linear form for the mapping, i.e.,  $g^{(f)}(y^{(f-1)}(\x)) = c \cdot y^{(f-1)}(\x)$, we recover the classic auto-regressive model~\citep{kennedy2000predicting}.
This method is further improved by \citet{le2013multi}, who used a deterministic parametric form of $g^{(f)}$ and an efficient numerical scheme to reduce the computational cost. 
Despite successfully dealing with some demonstrated problems, 
this parametric approach does not generalize well due to the difficulty of model selection and the demand for large training datasets~\citep{perdikaris2017nonlinear}.
To resolve this issue, a Bayesian non-parametric treatment can be implemented by placing a GP prior over the function $g^{(f)}$.
This is also known as the \textit{deep GP}~\citep{damianou2013deep}.
Although being capable of modeling complex problem, deep GP is infamous for its intractability. It requires computationally expensive variational approximation for model training and thus quickly becomes infeasible for multi-fidelity simulation problems.
\citet{perdikaris2017nonlinear} put forward a GP-based nonlinear autoregressive scheme with an additive structure,
\begin{equation} \label{eq:deep gp}
y^{(f)}(\x) = g^{(f)}\left(\x, y_*^{(f-1)}(\x)\right),
\end{equation}
where the error term is absorbed into $\g^{(f)}(\x)$, and
$\y_*^{(f-1)}(\x)$ is the posterior for $\x$ at fidelity $(f-1)$.
 This formulation specifies that $\y^{(f)}(\x)$ is a GP given the previous fidelity observations and the model inputs.
To better reflect the autoregressive nature, \citet{perdikaris2017nonlinear} suggested a covariance function that decomposes as
\begin{equation}
\label{eq:deep gp kernel}
\begin{aligned}
k^{(f)} & \left([\x, y_*^{(f-1)}(\x)], [\x',y_*^{(f-1)}(\x')]\right)\\ &=  k_x^{(f)}(\x, \x'; \btheta_{fx} ) \cdot k_y^{(f)}(y_*^{(f-1)}(\x), y_*^{(f-1)}(\x'); \btheta_{fy} ),
\end{aligned}
\end{equation}
where $k_x^{(f)}$ and $k_y^{(f)}$ are valid covariance functions with $\{\btheta_{f_x}, \btheta_{f_y}\}$ denoting the corresponding hyperparameters.
These hyperparameters could be learned via maximum-likelihood estimation as in Eq.~\eqref{eq:sgp likelihood} using input data $\{\X^{(f)}, \Y^{(f-1)}\}$ and output data $\Y^{(f)}$.
Thus, unlike the standard deep GP~\citep{damianou2013deep}, this autoregressive model does not have the intractable issues for estimating the unknown latent variables and hyperparameters.
This method requires estimation of $(l+3)$ hyperparameters with an ARD kernel for each fidelity model.

\subsection{Deep Coregionalization}
Despite being much more efficient than single-fidelity models, the GP autoregressive model does not naturally extend to multivariate cases, especially the high-dimensional ones. 
First, when using a common anisotropic kernel such as the ARD kernel, the deep kernel function $k_y^{(f)}$ in Eq.~\eqref{eq:deep gp kernel} requires $d$ hyperparameters to be optimized, which is practically impossible for our problem where $d$ can scale up to 1 million.
Second, it is unclear how to harness the strong output correlations. If we simply treat each output independently (given the inputs), we risk severe overfitting when learning with a small set of training examples, a common situation for the high-fidelity data.
\cmt{
\subsection{Deep Linear Model of Coregionalization}
In order to model the high dimensional output correlations within limited computational resources, we borrow the classic multiple output Gaussian process framework---Linear Model of Coregionalization (LMC)~\citep{matheron1982pour}---to decompose the $f$ fidelity high dimensional output,
\begin{equation} \label{eq:lmc}
\y^{(f)}(\x) = \sum_{r=1}^{R_f} \b^{(f)}_r z^{(f)}_r(\x) = \B^{(f)} \z^{(f)}(\x). 
\end{equation}
In this formulation, 
$\B^{(f)} = [\b^{(f)}_1,\dots,\b^{(f)}_{R_f}]$ is the collection of bases $\b^{(f)}_r \in \mathbb{R}^d$; $\z^{(f)}(\x) = [\z^{(f)}_1(\x),\dots,\z^{(f)}_{R_f}(\x)]^T$ is the collection of corresponding latent processes.
It is assumed that the latent processes are independent, i.e., $\cov( z^{(f)}_r(\x), z^{(f)}_{r'}(\x')) = 0$ for $r \neq r'$ and the bases are orthogonal.
The LMC model essentially assumes several independent latent processes to fully characterize the data variability. It is shown particularly efficient for the simulation-type problems\citep{higdon2008computer,zhe2019scalable}.
}
{
To model the output correlations, we can directly modify the model of~\citet{perdikaris2017nonlinear} by equipping Eq.~\eqref{eq:deep gp kernel} with an output correlation kernel as
\begin{equation}
\label{eq:deep gp kernel2}
\cov \left(y_{i}^{(f)}(\x), y_{j}^{(f)}(\x') \right)
=  k^{(f)}_d(i,j)  \cdot k_x^{(f)}(\x, \x' ) \cdot k_y^{(f)}(\y_*^{(f-1)}(\x), \y_*^{(f-1)}(\x')),
\end{equation}
Where $k^{(f)}_d(i,j)$ is an arbitrary valid kernel function encoding the output correlation between i-th output $y_{i}^{(f)}$ and j-th output $y_{j}^{(f)}$.
Directly working with Eq.~\eqref{eq:deep gp kernel2} is difficult as the output correlation can have maximum $d(d+1)/2$ hyperparameters. Fortunately, it has a much compact form.

\begin{lemma}
\label{LEMMA1}
If a multivariate GP's covariance function can be decomposed as Eq.~\eqref{eq:deep gp kernel2}, it must have an equivalent form of
\begin{equation}
\label{eq:icm}
\y^{(f)}(\x) 
= \b^{(f)} z^{(f)}\left(\x, \y_*^{(f-1)}(\x) \right),
\end{equation}
\sloppy where $z^{(f)}\left(\x, \y_*^{(f-1)}(\x) \right)$ is a GP with covariance function $k_x^{(f)}(\x, \x' ) \cdot k_y^{(f)}(\y_*^{(f-1)}(\x), \y_*^{(f-1)}(\x'))$ and $\b^{(f)} \in \mathbb{R}^{d\times1}$ is a base vector corresponding to the sum of eigenvectors times square roots of eigenvalues of kernel matrix $\K_d$, where $[K]_{ij}=k^{(f)}_d(i,j)$.
\end{lemma}

We leave the proof in Appendix~\ref{ap:proof_l1} for the sake of clarity.
The advantage of Eq.~\eqref{eq:icm} is that the number of (unknown) hyperparameters we need to consider reduces from $d(d+1)/2$ to $d$ for $f=1,\cdots,F$.
%
Substituting Eq.~\eqref{eq:icm} into Eq.~\eqref{eq:deep gp kernel2}, we have
\begin{equation}
\label{eq:icm sum gp}
\begin{aligned}
& \cov  \left(y_{i}^{(f)}(\x),y_{j}^{(f)}(\x') \right) \\
 & = k^{(f)}_d(i,j)  \cdot k_x^{(f)}(\x, \x' ) \cdot k_y^{(f)}(\b^{(f-1)} z_*^{(f-1)}(\x), \b^{(f-1)} z_*^{(f-1)}(\x')).
\end{aligned}
\end{equation}
This model is difficult to work with because $\b^{(f-1)} z_*^{(f-1)}(\x) \in \mathbb{R}^{d}$ and most practical kernels, \eg the ARD kernel, requires more than $d$ hyperparameters to be optimized.
However, note that $\b^{(f-1)} z_*^{(f-1)}(\x)$ is readily embedded in the subspace where $z_*^{(f-1)}(\x)$ lives. This particular structure allows us to find a compact representation of the kernel function $k_y^{(f)}(\b^{(f-1)} z_*^{(f-1)}(\x), \b^{(f-1)} z_*^{(f-1)}(\x'))$.
%
%
\begin{lemma}
\label{LEMMA2}
If a kernel function $k_y^{(f)}(\b^{(f-1)} z_*^{(f-1)}(\x), \b^{(f-1)} z_*^{(f-1)}(\x'))$ is stationary, \eg the ARD kernel, it must have a compact representation $k_z^{(f)}( z_*^{(f-1)}(\x), z_*^{(f-1)}(\x'))$, which is also a stationary kernel absorbing the base vector $\b^{(f-1)}$.
\end{lemma}

Lemma~\ref{LEMMA2}, whose proof is given in Appendix \ref{ap:proof l2}, suggests that the model complexity can be significantly reduced by absorbing the base vector $\b^{(f)}$ into the kernel function. This simplification makes modeling of very high-dimensional fields feasible. Applying this conclusion to Eq.~\eqref{eq:icm sum gp} gives a compact representation for the full kernel function: 
\begin{equation}
\label{eq:deep icm kernel}
\begin{aligned}
	\cov\left(y_{i}^{(f)}(\x),y_{j}^{(f)}(\x') \right) 
&= k^{(f)}_d(i,j) \cdot \hat{k}_z^{(f)}\left([\x,z_*^{(f-1)}(\x)], [\x',z_*^{(f-1)}(\x')]\right),\\
\end{aligned}
\end{equation}
where $\hat{k}_z^{(f)}\left([\x,z_*^{(f-1)}(\x)], [\x',z_*^{(f-1)}(\x')]\right)= k_x^{(f)}(\x, \x' ) \cdot k_z^{(f)}\left( z_*^{(f-1)}(\x), z_*^{(f-1)}(\x')\right) $ is a composite kernel of $k_x^{(f)}$ and $k_z^{(f)}$.
Rewriting Eq.~\eqref{eq:deep icm kernel} based on Lemma~\ref{LEMMA1}, we get
\begin{equation}
\label{eq:deep icm}
\y^{(f)}(\x) = \b^{(f)} \hat{z}^{(f)}(\x,z_*^{(f-1)}(\x)),
\end{equation}
where $\hat{z}_*^{(f-1)}(\x)$ is a GP with kernel function $ \hat{k}_z^{(f)}\left([\x,z_*^{(f-1)}(\x)], [\x',z_*^{(f-1)}(\x')]\right)$.
Recursively using Eq.~\eqref{eq:deep icm} and Lemma~\ref{LEMMA2} for $f=1,\cdots,F$, we can see clearly that the lower-fidelity knowledge propagates to the higher-fidelity model only through the univariate latent process $\hat{z}_*^{(f-1)}(\x)$, which can be solved efficiently using univariate autoregression model of \citet{perdikaris2017nonlinear}.

\begin{remark}
\label{THEOREM1}
If each multi-fidelity high-dimensional autoregressive model has a covariance structure of Eq.~\eqref{eq:deep gp kernel2}, the high-fidelity function $\y^{(F)}(\x)$ dependents only on an univariate autoregressive model and a base vector.
\end{remark}

Remark~\ref{THEOREM1} implies that the propagation of the fidelity correlations in a high-dimensional model can be achieved equivalently thought the propagation of latent processes, which are independent from the output correlations.
%
However, the model suggested by Remark~\ref{THEOREM1} is restrictive and can not generalize well to a wide variety of problems. 
First, the kernel function is assumed to be stationary, which greatly limits the capacity of a GP model.
Second, Eq.~\eqref{eq:deep gp kernel2} assumes an over simplified kernel structure.
Comparing Lemma~\ref{LEMMA1} with the LMC of Eqs.~\eqref{eq:lmc} and \eqref{eq:lmc in sum},
we immediately note that Lemma~\ref{LEMMA1} corresponds to a special case of
LMC, where only one latent process and one base vector composed via eigen analysis of a simple output correlation matrix is considered.
As suggested in \citet{higdon2008computer}, a complicated stochastic process, even a non-separable one, can be well approximated using a sum of simple latent processes.
Thus, we place a prior of GP sum for the spatial-temporal field of each fidelity,
\begin{equation}
\label{eq:lmc multi}
\y^{(f)}(\x) = \sum_{r=1}^{R_f} \b^{(f)}_r z^{(f)}_r(\x) = \B^{(f)} \z^{(f)}(\x),
\end{equation}
where $R_f$ is the number of latent processes, $\{\b^{(f)}_r\}$ are orthogonal base vectors, \ie ${\B^{(f)}}^T \B^{(f)} = \I$, and $\{z^{(f)}\}_{r=1}^{R_f}$ are independent latent processes (GPs). 
This formulation is consistent with the LMC of Eq.~\eqref{eq:lmc}, which has shown to be general and applicable for decomposing spatial-temporal fields $\y^{(f)}(\x)$ from a simulator~\citep{higdon2008computer,zhe2019scalable}. 
The formulation of Eq.~\eqref{eq:lmc multi} represents a generalization of covariance structure suggested by Eq.~\eqref{eq:deep gp kernel2} using a additive structure 
\begin{equation}
\cov \left(y_{i}^{(f)}(\x), y_{j}^{(f)}(\x') \right)
=  \sum_{r=1}^{R_f} k^{(f)}_{dr}(i,j)  \cdot k_{xr}^{(f)}(\x, \x' ) \cdot k_{yr}^{(f)}(\y_*^{(f-1)}(\x), \y_*^{(f-1)}(\x')),
\end{equation}
where $k^{(f)}_{dr}, k_{xr}^{(f)}$ and $k_{yr}^{(f)}$ correspond to the $r$-th latent processes of between-output correlations, between-input correlations, and cross-fidelity correlation, respectively. 
$k^{(f)}_{dr}$ is simplified to have maximum $d$ hyperparameters.
To relax the stationary assumption made by Lemma~\ref{LEMMA2}, we construct the deep coregionalization model on the general autoregressive model of Eq.~\eqref{eq:deep gp}.
%
%
%
%
%
\cmt{
------\\	
	which gives rise to a covariance matrix between high-dimensional outputs,
	\begin{equation}
	\label{eq:kernel separable}
	\begin{aligned}
	\cov\left(\y^{(f)}(\x),\y^{(f)}(\x') \right)
	=  \K^{(f)}_d   \cdot k_x^{(f)}(\x, \x' ) \cdot k_y^{(f)}(y_*^{(f-1)}(\x), y_*^{(f-1)}(\x')) + \tau^{-1},
	\end{aligned}
	\end{equation}
	where $[\K^{(f)}_d]_{ij} = k^{(f)}_d(i,j)$ is the output correlation matrix for fidelity-$f$.
	
	Lemma: if a covariance matrix has the form of Eq. 
	
	Assuming that Eq.\eqref{eq:kernel separable} applies to all fidelities and $k_y^{(0)}=1$, \ie no lower-fidelity information is taken for $\y^{(1)}(\x)$, 

We immediately recognize this is a simplified case of Eq.~\eqref{eq:lmc in sum}, where $\b_r = \v^{(f)}_r \sqrt{\lambda^{(f)}_r}$ and $k_r(\x,\x') =  k_x^{(f)}(\x, \x' ) \cdot k_y^{(f)}(y_*^{(f-1)}(\x) y_*^{(f-1)}(\x'))$. $\v_r$ and${\lambda_r}$ are the eigenvector and eigenvalue of $\K^{(f)}_d$ here.
Thus, we can write $\y^{(f)}(\x)= \B^{(f)}z(\x) $ as in Eq.\eqref{eq:lmc}.
Similarly, 

With some linear algebra (details are supplied in Appendix XX), we can show that the separable structure is equivalent to specify that 
\begin{equation}
\y^{(f)}(\x) = \v k_y^{(f)}(y_*^{(f-1)}(\x), y_*^{(f-1)}(\x')),
\end{equation}
------\\
}
%
%
%
%
Substituting Eq.~\eqref{eq:lmc multi} into Eq.~\eqref{eq:deep gp}, we have
\begin{equation}
\begin{aligned}
\B^{(f)} \z^{(f)}(\x) & = \g^{(f)}\left(\x, \B^{(f-1)} \z_*^{(f-1)}(\x)\right)\\
\z^{(f)}(\x) & = {\B^{(f)}}^T \g^{(f)}\left(\x, \B^{(f-1)} \z_*^{(f-1)}(\x)\right)\\
\z^{(f)}(\x) & = \hat{\g}^{(f)}\left(\x,\z_*^{(f-1)}(\x)\right),\\
\end{aligned}
\label{eq:deep lmc1}
\end{equation}
where $\hat{\g}^{(f)}$ is the composite function absorbing $\B^{(f)}$ and $\B^{(f-1)}$ into the multivariate mapping $\g^{(f)}$.
%
%
\begin{theorem}
	\label{THEOREM2}
	If each fidelity output admit an LMC formulation (as Eq.~\eqref{eq:lmc multi}), $\hat{\g}_r^{(f)}$ must exist and can be decomposed into several independent univariate GPs, $\{\hat{g}_r^{(f)}(\x,\z_*^{(f-1)}(\x))\}_{r=1}^{R_f}$.
\end{theorem}
We provide the proof in Appendix \ref{ap:proof t2}.
With this conclusion, we can see that the high-fidelity field depends only on the latent process of its previous fidelity and the inputs once the bases are found. The output correlations are captured independently for each fidelity, whereas the fidelity correlations are propagated through independent univariate GPs. We thus call the model \textit{deep coregionalization}.
The joint likelihood given the multi-fidelity data is
\begin{equation} 
\label{eq:dc likelihood}
\Lcal \propto \sum_{f=1}^{F}  \sum_{r=1}^{R_f} \frac{1}{2}  \ln|\bSigma_{fr}| - \frac{1}{2}\tr\left({\z^{(f)}_r}^T \bSigma_{fr}^{-1} \z^{(f)}_r \right),
\end{equation}
where $\z^{(f)}_r = \Y^{(f)}\b_r^{(f)}$ is the projection of $\Y^{(f)}$ onto base $\b_r^{(f)}$, $\bSigma_{fr} = \K_{fr} + \tau_{fr}^{-1} \I$ is the covariance matrix with the noise term for r-th latent process of f-fidelity observations, 
and $[\K_{fr}]_{ij} = k_{fr}\left(\x_i,\z^{(f-1)}(\x_i),\x_j,\z^{(f-1)}(\x_j) \right)$ is the covariance matrix for the latent process $\hat{\g}_r^{(f)}(\x,\z_*^{(f-1)}(\x))$ based on its own kernel function $k_{fr}$ and inputs $\X^{(f)}$ and $\Z^{(f-1)}_{\e_f,:}$

\subsection{Residual Principal Component Analysis}
The challenge for \ours is the optimization of the bases $\B^{(f)}$ and the low-rank $R_f$ for each fidelity.
If we set $R_f=1$ for each fidelity data, we recover the \textit{intrinsic coregionalization model} (ICM)~\citep{goovaerts1997geostatistics} where an efficient Kronecker structure could be utilized to speed up the computation. However, the model capacity can be quite limited due to the oversimplified latent process. A larger $R_f$ enables us to model outputs with different characteristics~\citep{alvarez2012kernels} with the cost of more hyperparameters to infer.
The output correlation can be indirectly modeled using a full-rank or low-rank Cholesky decomposition to allow flexible output correlations~\citep{bonilla2008multi}.
The number of parameters to infer for the aforementioned methods is proportional to $d \times R_f$, resulting into a impractical approach for high-dimensional outputs.  
Since the simulation outputs are always complete and can be efficiently represented via principal components~\citep{ramsey1997functional},
\citet{higdon2008computer} suggested using the principal components as the bases. 
More specifically, after standardizing the data, singular value decomposition (SVD) is applied to $\Y^{(f)}$, and the first $R_f$ left singular vectors are then used as the bases $\B^{(f)}$.
Another benefit of this approach is a heuristic method to choose the number of bases by choosing $R_f$ such that at least $90\%$ of the variance of the output is covered by the bases. 
It is reported that the components that explain minor trend of the variation do not add to the predictive ability of the GP model~\citep{higdon2008computer}.

This approach works well for many single-fidelity applications~\citep{higdon2008computer,zhe2019scalable}, but a direct implementation to our model here is inappropriate.
First, the principal components (the bases) are essentially based on the empirical covariance matrix given data $\Y^{(f)}$. 
Since we can afford to have only a limited number of high-fidelity samples, the empirical covariance matrix is a very rough approximation to the real one, and the resulting principal components are consequently inaccurate.
Second, even provided that we are always given enough samples to explore the output correlation at each fidelity, direct implementation of \citet{higdon2008computer} is of low efficiency.
According to our experiments, when directly implementing SVD, the dominant components and their corresponding coefficients for each fidelity are always similar.
The similarity implies that a using a complicated nonlinear model, \eg GP, to propagate such a simple correlation in a deep structure is of low efficiency and may lead to inferior model accuracy.
Instead, we can limit the information passing down the model and only model the updated (added) information.
Specifically, we specify
\begin{equation}
\begin{cases}
\y^{(f)}(\x) = \B^{(f)} \z^{(f)}(\x) \;\; &\mathrm{for} \; f = 1 \\
\y^{(f)}(\x) = \B^{(f)} \z^{(f)}(\x) + \y^{(f-1)}(\x)  \;\; &\mathrm{for} \; f \geq 2. \\
\end{cases}
\end{equation}
In this formulation, each fidelity layer learns the residual (additional) information compared with the previous fidelity layer rather than the whole output information, and thus we call this a \textit{residual deep structure} similar to the work of \citet{he2016deep}.
Except for efficiency, another advantages of this approach is that the predictive posterior for high-fidelity results naturally decompose into an additive structure.
The additive structure can help us identify the main sources of uncertainty and adjust our model at each fidelity accordingly to reduce model uncertainty and improve model accuracy.
%

Inspired by the practical approach of \citep{higdon2008computer} to extract the bases, we introduce the \textit{residual principal component analysis} (ResPCA) for our model. 
The method is a modification of PCA applied to the residual information instead of the original data. We use SVD, which is more numerical stable than direct eigendecomposition on the empirical covariance matrix, to extract the eigenvectors. The detailed steps of this method are presented in Algorithm~\ref{algo:rpca}.
\begin{algorithm}[h]
	\caption{ResPCA}
	\label{algo:rpca}
	\begin{algorithmic}[1]
		\Require
		multi-fidelity multivariate data $\{\X^{(f)},\Y^{(f)} \}_{f=1}^F$ and index $\{\e_{f}\}_{f=1}^F$;
		\Ensure
		Residual bases $\{\B^{(f)}\}_{f=1}^F$;
		Residual coefficients $\{\Z^{(f)}\}_{f=1}^F$;
		\State $\B^{(1)} = \mathrm{SVD}({\Y}^{(1)})$; \; $\Z^{(1)} = {\Y}^{(1)} {\B^{(1)}}^T$
		\For{each $ f \in [2,F]$}
		\State Achieve residual information $\widetilde{\Y}^{(f)} = \Y^{(f)} - \Y^{(f-1)}_{\e_{f},:} $
		\State Achieve residual base, $\B^{(f)} = \mathrm{SVD}(\widetilde{\Y}^{(f)})$
		\State Achieve residual coefficients, $\Z^{(f)} =  {\B^{(f)}}^T \widetilde{\Y}^{(f)}$
		\EndFor
		\label{code:recentEnd}
	\end{algorithmic}
\end{algorithm}
We can now substitute the projections of residual information $\Z^{(f)}$ into Eq.~\ref{eq:dc likelihood} for model training.

\subsection{Prediction and Uncertainty Propagation}
Due to the intractability of a deep GP structure, except for the latent posterior $\z^{(1)}(\x)$ corresponding to the lowest fidelity $\y^{(1)}(\x)$, the predictive latent posterior for each fidelity is
\begin{equation}
\label{eq:dgp posterior}
\z_*^{(f)}(\x) = \int \g^{(f)}\left(\x, \z_*^{(f-1)}(\x) \right) p\left(\z_*^{(f-1)}(\x)\right) d \z_*^{(f-1)}(\x).
\end{equation}
This integral is intractable and the posterior is no longer a Gaussian, but we can approximate it numerically using sampling-based methods. We can further reduce the complexity due to the independent assumption of the LMC; Eq.~\eqref{eq:dgp posterior} naturally decomposes as
\begin{equation}
\label{eq:dgp posterior decomp}
\z_{*}^{(f)}(\x) =  \prod_{k=1}^{R_f} \int g_k^{(f)}\left(\x, \z_*^{(f-1)}(\x) \right) \prod_{r=1}^{R_{f-1}} p(z_{*r}^{(f-1)}(\x)) d z_{*r}^{(f-1)}(\x).
\end{equation}
We can now easily implement a Monte Carlo sampling method to calculate the latent process posteriors. Another way to solve the intractable integral is to apply a Gaussian approximation of each latent posterior as in \citet{perdikaris2017nonlinear} and \citet{girard2003gaussian}.
Once we obtain the approximated posterior, the first two order statistic admits a tractable solution due to the summation structure.
Given the posterior of the latent processes $\z^{(f)}(\x)$ and the bases $\B^{(f)}$, the mean and variance of the finest fidelity field $\y^{(F)}(\x)$ can be calculated as
\begin{equation}
\begin{aligned}
\EE[\y^{(F)}(\x)] &= \sum_{f=1}^F \B^{(f)}\EE[\z_*^{(f)}(\x)],\\
\var[\y^{(F)}(\x)] &= 
\sum_{f=1}^F \left( \B^{(f)}\var[\z_*^{(f)}(\x)]{\B^{(f)}}^T \right)\\
&- 2 \sum_{f\neq f'} \left(\B^{(f)}\cov[\z_*^{(f)}(\x),\z_*^{(f')}(\x)]{\B^{(f')}}^T \right),
\end{aligned}
\label{eq:y post}
\end{equation}
where $\var[\z^{(f)}(\x)]$ is the diagonal covariance matrix of the fidelity-$f$ latent posterior, and $\cov[\z^{(f)}(\x),\z^{(f')}(\x)$ is the covariance matrix of $\z^{(f)}(\x)$ and $\z^{(f')}(\x)$. This covariance matrix does not have an analytical form and 
can be calculated empirically using sampling methods. 
 

\subsection{Implementation and Model Complexity}
Before we describe our experiments, we review our model and the its implementation details for the emulation of spatial-temporal fields.
1). To build a surrogate model, We need to firstly collect different fidelity data from a simulator at design inputs $\X$. To better explore the response surface, we use a Sobol sequence~\citep{sobol1976uniformly} to generate the design points. For fidelity-$f$, we choose the first $N_f$ design points of $\X$ to generate the corresponding outputs $\Y^{(f)}$. The computational cost at this stage depends on the simulator and the fidelity setting.
2). We then apply Algorithm~\ref{algo:rpca} to get the residual bases $\{\B^{(f)}\}_{f=1}^F$ and corresponding projections $\{\Z^{(f)}\}_{f=1}^F$. The computational cost for fidelity-$f$ data is $\Ocal(\min(N_f^2 d,N_f d^2) )$ using SVD. For the choice of low-rank $R_f$, we can set a variance ratio as in \citet{higdon2008computer}. Specifically, the minimum number of bases to capture the variance ratio (calculated by the eigenvalues of the covariance matrix) is used.
3). We maximize the joint likelihood function of Eq.~\eqref{eq:dc likelihood} w.r.t. the hyperparameters using gradient-based optimization method, \eg L-BFGS-B. 
This optimization consists of $\sum_{f=2}^{F}(R_f(R_{f-1}+3))$ hyperparameters and the computational complexity is $\Ocal(R_f N_f^3)$ per iteration.
In cases where we have very limited training data, we can instead place a conditional independent assumption~\citep{Lawrence06GPLVM} for $\hat{g}^{(f)}_r(\x,\z^{(f-1)}_*(\x))$, which reduces the hyperparameters to $\sum_{f=2}^{F} (R_{f-1}+3)$ and complexity $\Ocal(N_f^3)$ compared to the fully independent approach.
For large numbers of observations, sparse GPs~\citep{snelson2005sparse} can be placed on the latent process to reduce the computational cost.
4). Given a new $\x_*$, we calculate each latent posterior using Eq.~\eqref{eq:dgp posterior} and then compute the mean and variance of the highest fidelity outputs using Eq.~\eqref{eq:y post}.



\section{Experiments}\label{sect:expr}
In this section, we first examine our model with three canonical PDEs and compare it with other methods in terms of model capacity and accuracy.
We then apply our model to a fluid dynamic problem to demonstrate the advantages of our model in more complicated real-world applications.

\noindent\textbf{Competing methods}. We compared \ours (DC) with four low-rank GP models for emulations of spatial-temporal fields:  
(1) PCA-GP~\citep{higdon2008computer}, the popular LMC-based GP model for high-dimensional simulation data via principal component analysis (PCA), 
(2) IsoMap-GP~\citep{xing2015reduced}, an extension of PCA-GP based on IsoMap~\citep{balasubramanian2002isomap}, a classic nonlinear dimension reduction method,
(3) KPCA-GP~\citep{xing2016manifold}, another extension of PCA-GP based on implicit nonlinear bases through kernel PCA~\citep{scholkopf1998nonlinear},
and (4) HOGP~\citep{zhe2019scalable}, a very recent approach that tensorizes the outputs and introduces latent coordinate features (in tensor space) to model the output correlations.  

\noindent\textbf{Parameter settings}.
All models were implemented in Matlab using L-BFGS-B with a maximum of 100 iterations for model training and using ARD for all kernel functions.
For \ours, a single low-rank $R$ is equally applied to each fidelity.
IsoMap-GP and KPCA-GP used $10\%$ of the training data to solve the additional pre-image problems, as is suggested in \citet{xing2015reduced,xing2016manifold}.
%
%
For each dataset, \ours integrates the examples of all the fidelities for training. 
Since the other methods work only on single-fidelity data, we conducted their training based on the examples of each fidelity separately. 
For instance, PCA-GP-F1 denotes PCA-GP trained with samples of fidelity-1, whereas PCA-GP-F2 with fidelity-2. 

\noindent\textbf{Evaluation}.
We varied the number of bases and the number of training samples to compare the performance of each method.
%
The performance was measured using the root-mean-square error (RMSE), defined as
$RMSE = \sqrt{\sum_{i,j}(\hat{y}_{ij}-y_{ij})^2/Nd },$
where here $\hat{y}_{ij}$ is the j-th dimension of prediction $\hat{\y}_{i}$, $\y_i$ is the ground truth, and $i=1,\ldots,N$ is the index of the total $N$ test points.
We also used the mean absolute error (MAE) field, defined as $\sum_{i}(|\hat{\y}_{i}-\y_{i}|)/N $, to demonstrate the local error.
Data was normalized beforehand to provide a fair comparison between all competing methods.

\subsection{Modeling Fundamental PDEs}
We consider three canonical PDEs: Poisson's equation, the heat equations, and Burger's equation.
These PDEs play important roles in scientific and engineering applications~\citep{chapra2010numerical,chung2010computational,burdzy2004heat}. 
They provide some common scenarios in simulations, such as high-dimensional spatial-temporal field outputs, nonlinearities, and discontinuities, and are often used as benchmark problems for surrogate models~\citep{lagaris1998artificial,tuo2014surrogate,efe2003proper,raissi2017machine}. 
For our experiments, Poisson's equation, $\Delta u = 0$ considered a spatial domain $\textbf{x} \in [0,1] \times [0,1]$ with Dirichlet boundary conditions. The simulator was parameterized by the initial condition of the four boundaries and the center of the rectangle domain from $0.1$ to $0.9$ and was solved using finite difference.
%
The 1D viscous Burger's equation, $\frac{\partial u}{\partial t} + u \frac{\partial u}{\partial x} = v \frac{\partial^2 u}{\partial x^2},$ where $u$ represents the volume, $x$ indicates a spatial location, $t$ denotes the time, and $v$ represents the viscosity, considered solutions for  $x\in[0,1]$, $t \in [0,3]$ with a initial condition of $u(x,t_0)=\sin(x\pi/2)$ and homogeneous Dirichlet boundary conditions. The simulator was parameterized by the viscosities $v \in [0.001,0.1]$ and solved using the finite elements method. 
The 1D heat equation, $\frac{\partial u}{\partial t} + \alpha \Delta u =0 $, where $u$ represents the heat, $\alpha$ the thermal conductivity, and $\Delta$ the Laplace operator, considered the solutions for $x\in[0,1]$ and $t \in [0,5]$ with Neumann boundary conditions and initial conditions of $u(x,0)=H(x-0.25)-H(x-0.75)$, where $H(\cdot)$ is the Heaviside step function. The simulator is parameterized by the flux rate of the left boundary at $x=0$ (ranging from 0 to 1), the flux rate of the right boundary at $x=1$ (ranging from -1 to 0), and the thermal conductivity (ranging from 0.01 to 0.1); it was solved using finite difference in space and backward Euler in time.


\noindent \textbf{Simulation and data generation.}
The number of node/steps used in the solver determines the data fidelity: the more nodes/steps in the solver, the higher the fidelity of the results.
For each PDE, we first generated $256$ design inputs using Sobol sequence~\citep{sobol1976uniformly} and $128$ test inputs using uniform sampling.
For each input, three fidelity field outputs are generated by running the simulation with three meshes, \ie $16\times16$, $32\times32$, and $64\times64$ regular rectangular meshes.
The simulation results are recorded using a $100\times100$ spatial-temporal (or just spatial) grid. 
Since random shuffling can invalidate the Sobol sequence for a better response surface coverage, we did not test each method using cross-validation.

We first conducted a two-fidelity test.
We provided each model with $256$ fidelity-1 samples and varied the number of fidelity-2 samples from $\{8,16,32,64, 128, 256\}$ and the model low-rank $R$ from $\{4,8,16\}$. 
The fidelity-2 test points were used as ground truths for validating the predictive results.
The RMSE as a function of fidelity-2 training samples is shown in Fig.~\ref{fig:rmse 2lv}. 
As we can see, \ours obtains the smallest prediction error in most cases except for the heat equation with $R=16$ and 16 fidelity-2 training samples.
This exception may be caused by using too many latent processes (high $R$) in the deep structure, which makes model training difficult with limited data.
As we have more fidelity-2 samples, this fault disappears.
We should emphasize that with only $16$ fidelity-2 samples, \ours can outperform most other methods with 6 times more, \ie $256$, fidelity-2 samples.

To demonstrate the detailed error, we show the MAE fields using $256$ fidelity-1 samples and $64$ fidelity-2 samples 
with model low-rank $R=8$ in Fig.~\ref{fig:local 2lv}. Six actual predictions compared with the ground truth fields for all methods are also included in Appendix \ref{app:actual predictions lv2}.
Note that methods solely relying on fidelity-1 data often produce significant errors. 
The performances of the Fidelity-2-based methods are overall better than fidelity-1-based methods, but there are exceptions, such as HOGP-F2 and ISOMAP-GP-F2 for Poisson's equation. In contrast, \ours shows consistent minimal MAE fields among all competing methods.

\begin{figure}[htbp]
	\centering
	\begin{subfigure}[t]{0.8\textwidth}
		\centering
		\includegraphics[width=\textwidth]{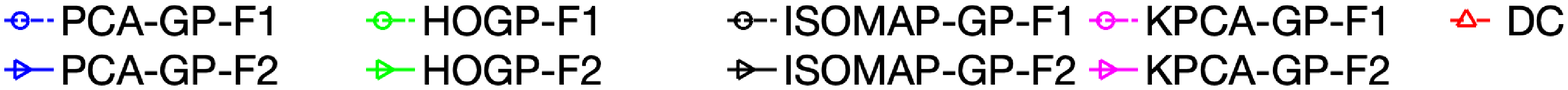}
		\vspace{-0.1in}
	\end{subfigure} \\
	\begin{subfigure}[t]{0.9\textwidth}
		\centering
		\includegraphics[width=\textwidth]{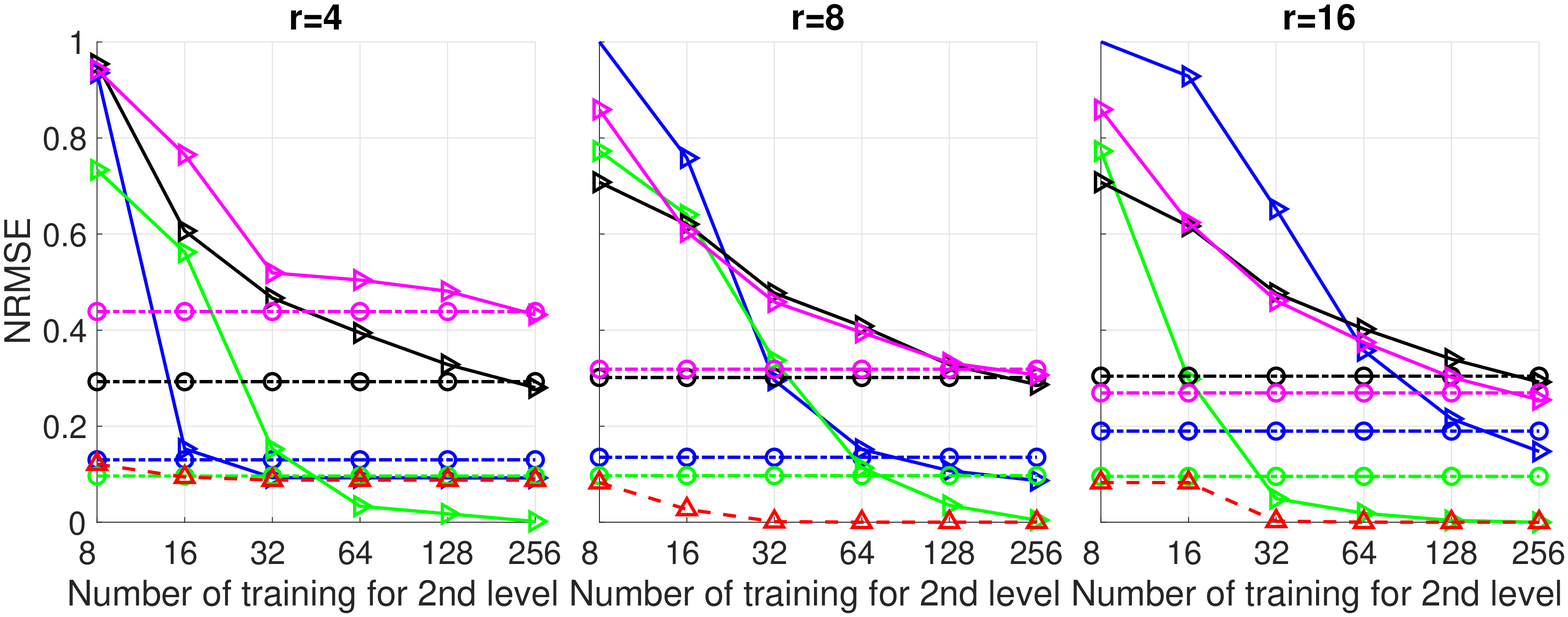}
	\end{subfigure} \\
	\begin{subfigure}[t]{0.9\textwidth}
		\centering
		\includegraphics[width=\textwidth]{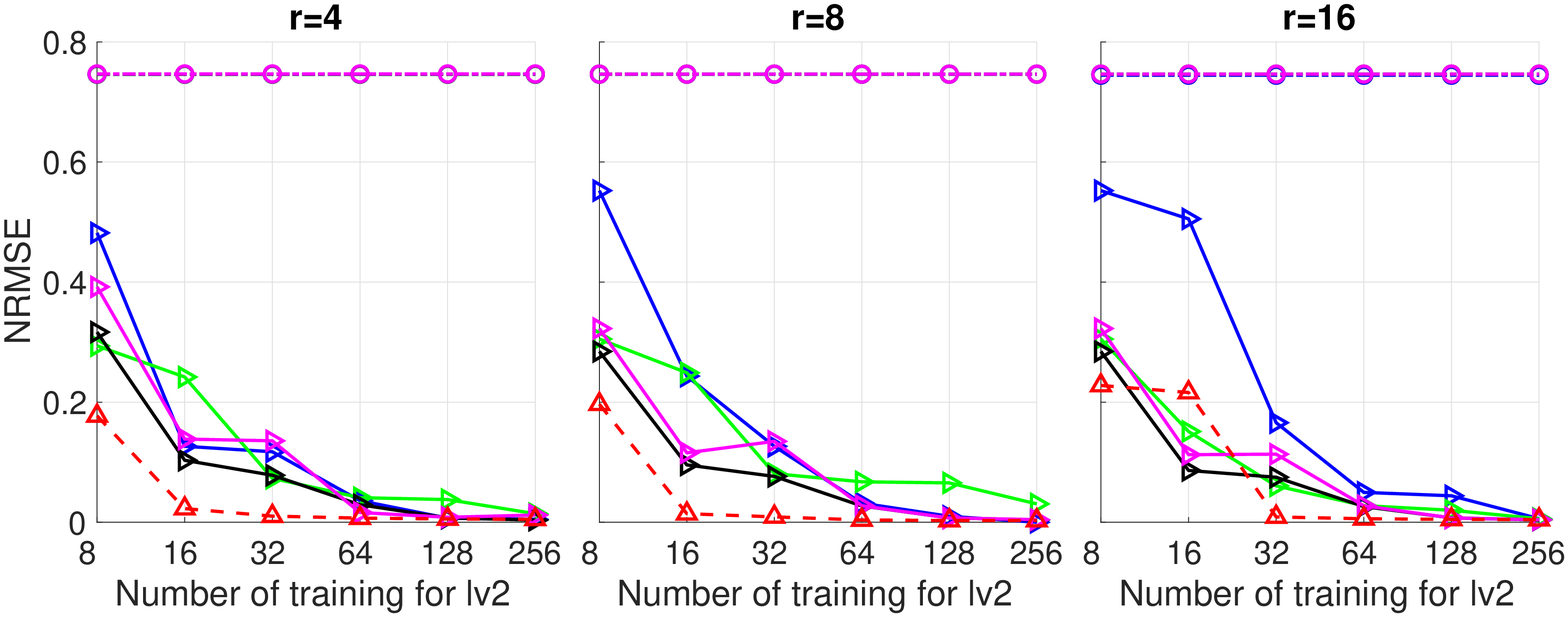}
	\end{subfigure} \\
	\begin{subfigure}[t]{0.9\textwidth}
		\centering
		\includegraphics[width=\textwidth]{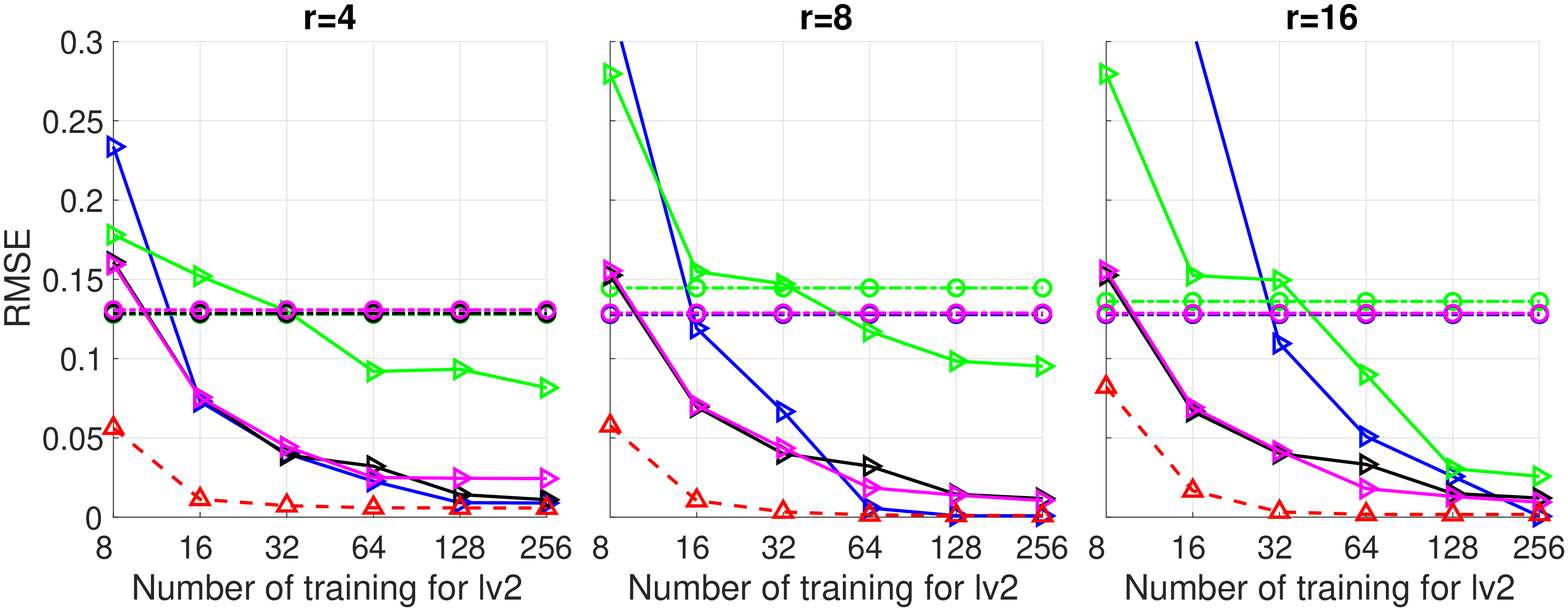}
	\end{subfigure} \\	
	\caption{RMSE for Poisson's equation (top row), the heat equation (middle row) and Burger's equation (bottom row) using training data from 2 fidelities. The number of fidelity-2 samples varies from 8 to 256; the number of fidelity-1 samples is fixed to 256; the low-rank $R$ varies from $\{4,8,16\}$.}
	\label{fig:rmse 2lv}
\end{figure} 
\begin{figure}[htbp]
	\centering
	\begin{subfigure}[t]{1\textwidth}
		\centering
		\includegraphics[width=\textwidth]{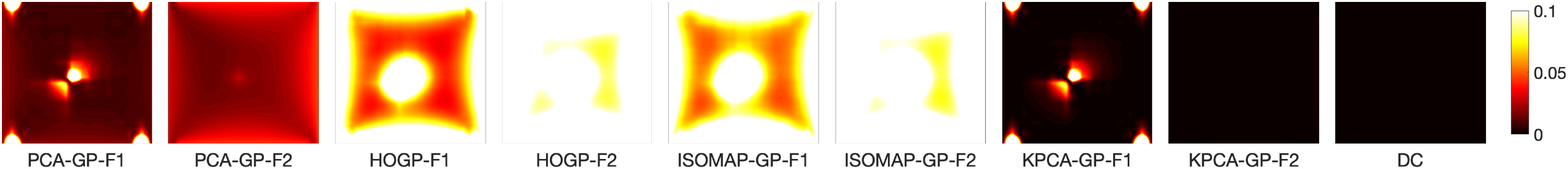}
	\end{subfigure} \\
	\begin{subfigure}[t]{1\textwidth}
		\centering
		\includegraphics[width=\textwidth]{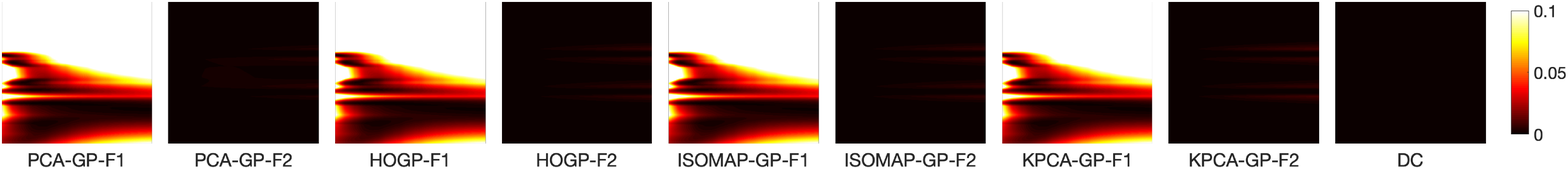}
	\end{subfigure} \\
	\begin{subfigure}[t]{1\textwidth}
		\centering
		\includegraphics[width=\textwidth]{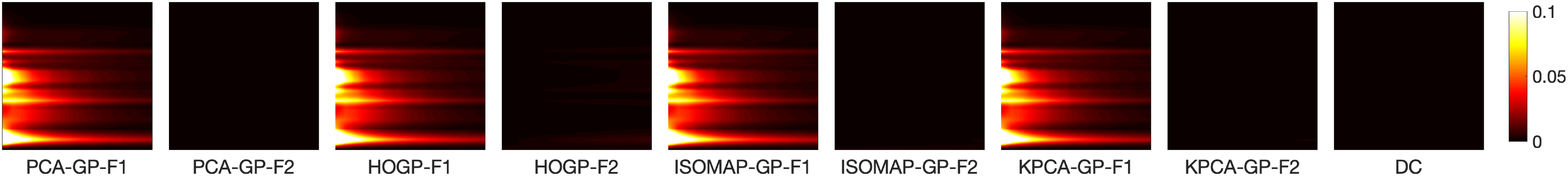}
	\end{subfigure} \\
	\caption{MAE field for Poisson's equation (top row), the heat equation (middle row) and Burger's equation (bottom row) using 256 fidelity-1 samples and 64 fidelity-2 samples with row-rank $R=8$}
	\label{fig:local 2lv}
\end{figure} 

Predicting the fidelity-2 can be easy considering how simple the fidelity-2 simulations are. Thus, we extend the experiment to include three fidelities and to predict the $128$ fidelity-3 fields.
Since it is difficult to show all the combinations of different training number settings for three fidelities, we introduced the fidelity ratio, defined by the ratio of training points at different fidelities.
We can change the fidelity ratio freely here to explore the influence of the training setting to the model performance.  
In practice, this fidelity ratio should be adjusted to reflect the ratio of simulation costs to maximize the efficiency.
The RMSEs of fidelity ratios of $4:2:1$ and $16:4:1$ are shown in Figs.~\ref{fig:rmse 3lv_2} and \ref{fig:rmse 3lv_1}, respectively. 
The x-axis denotes the number of training samples at fidelity-1 (and consequently samples at fidelity-2 and fidelity-3).
Overall, the fidelity-1 and fidelity-2 based methods show limited improvements given more training data since the low fidelity data do not contain high-fidelity information for the models to learn. 
The fidelity ratio has a huge impact on the fidelity-3-based methods because these method rely merely on the fidelity-3 samples; with only a few fidelity-3 samples, they naturally perform badly.
In contrast, the fidelity ratio has a lower influence on the performance of \ours; as the training samples increase, the performance converge to a similar level, which is also better than other methods with even $256$ fidelity-3 samples. 
Except for being stable, \ours outperforms other methods with a large margin at the same setting in most cases, especially with the fidelity ratio of $4:2:1$.


Similar to the previous experiments, the MAE fields of models using 64, 16, and 4 training samples (for fidelity-1, -2, and -3) with low-rank $R=8$ are shown in Fig.~\ref{fig:local 3lv_1}; six actual predictions along with the corresponding ground truth fields are included in Appendix \ref{app:actual predictions lv3}.
At this setting, the fidelity-3 based methods are no better than the fidelity-2 based method due to the lack of training samples.
Despite lacking the high-fidelity information, the fidelity-1 based methods  can sometimes outperform their counterparts based on higher fidelity data, such as PCA-GP-F1 and KPCA-GP-F1 for Poisson's equation.
Overall, the performance of each model varies from problem to problem except for \ours, which consistently shows the best performance for all cases here.

\begin{figure}[htbp]
	\centering
	\begin{subfigure}[t]{0.8\textwidth}
		\centering
		\includegraphics[width=\textwidth]{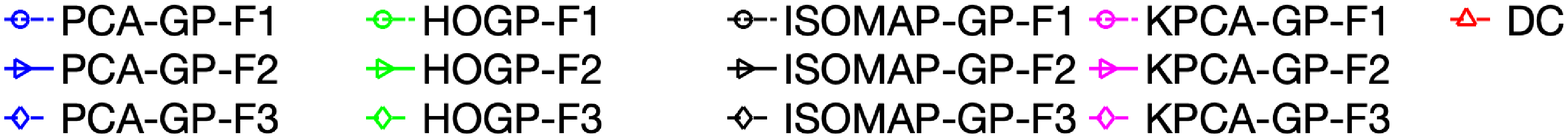}
		\vspace{-0.1in}
	\end{subfigure} \\
	\begin{subfigure}[t]{0.9\textwidth}
		\centering
		\includegraphics[width=\textwidth]{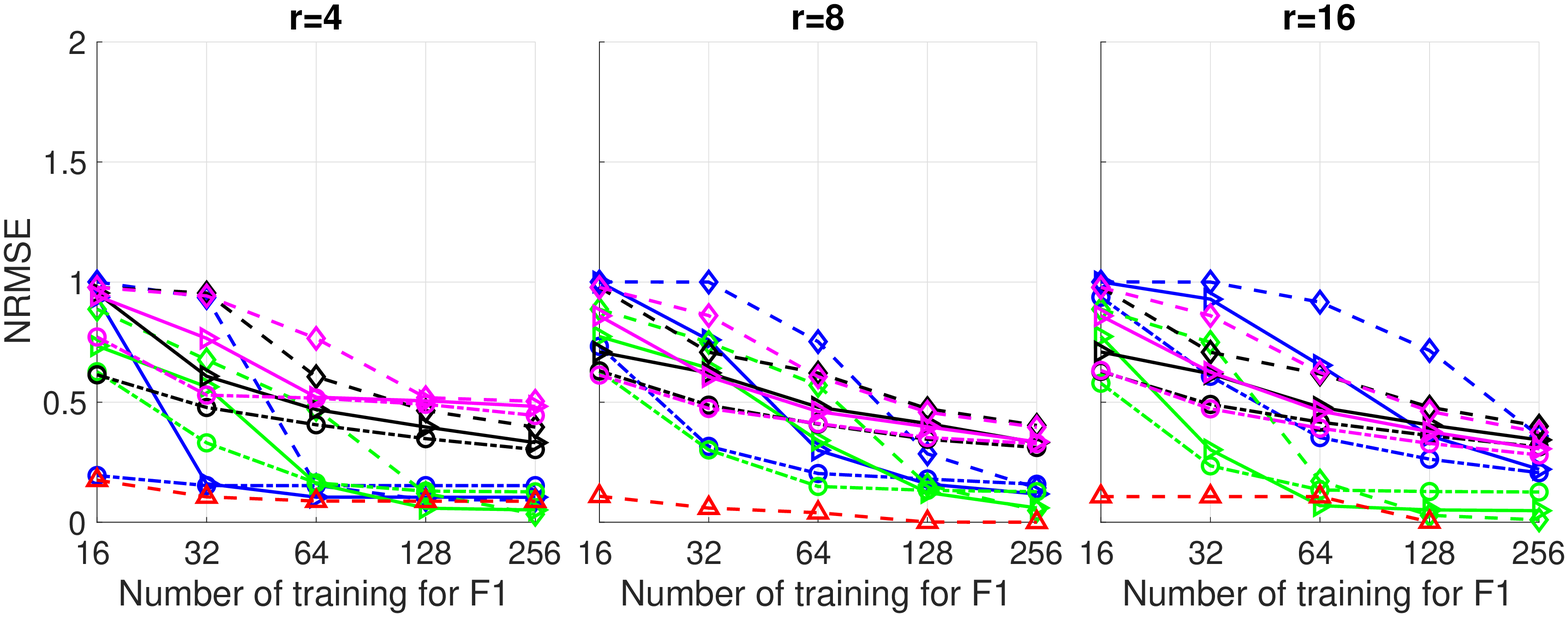}
	\end{subfigure} \\
	\begin{subfigure}[t]{0.9\textwidth}
		\centering
		\includegraphics[ width=\textwidth]{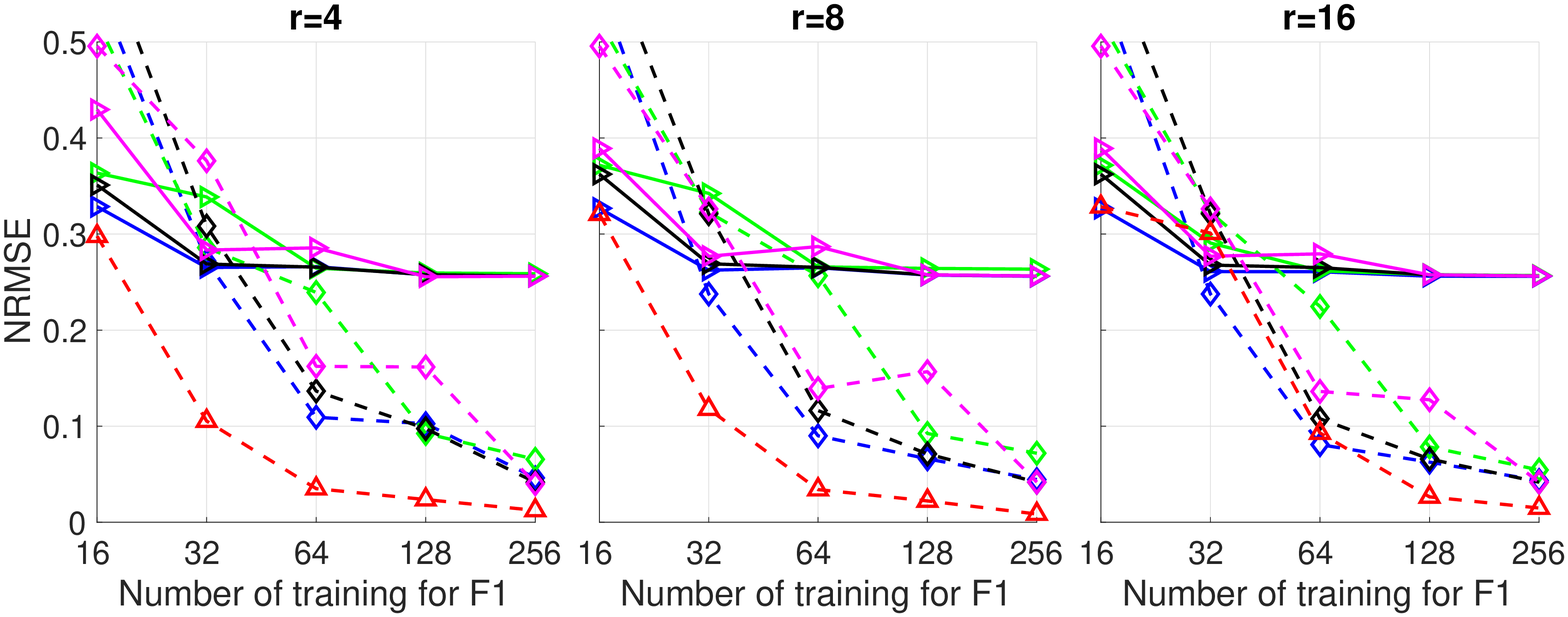}
	\end{subfigure} \\
	\begin{subfigure}[t]{0.9\textwidth}
		\centering
		\includegraphics[ width=\textwidth]{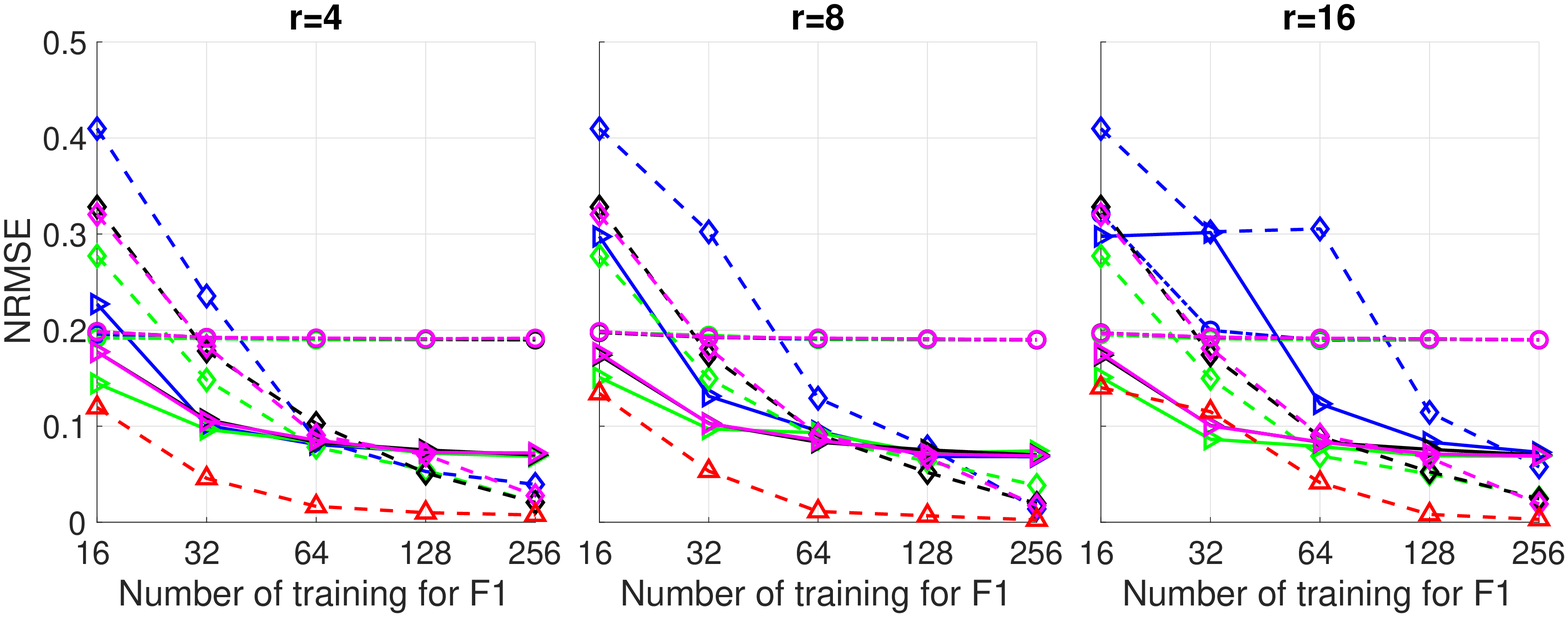}
	\end{subfigure} \\	
	\caption{RMSE for Poisson's equation (top row), the heat equation (middle row) and Burger's equation (bottom row) with data of 3 fidelity, low-rank $R=\{4,8,16\}$, and fidelity ratio of $4:2:1$.}
	\label{fig:rmse 3lv_2}
\end{figure}

\begin{figure}[htbp]
	\centering
	\begin{subfigure}[t]{0.8\textwidth}
		\centering
		\includegraphics[width=\textwidth]{./fig_v2/legendx3.eps}
		\vspace{-0.1in}
	\end{subfigure} \\
	\begin{subfigure}[t]{0.9\textwidth}
		\centering
		\includegraphics[width=\textwidth]{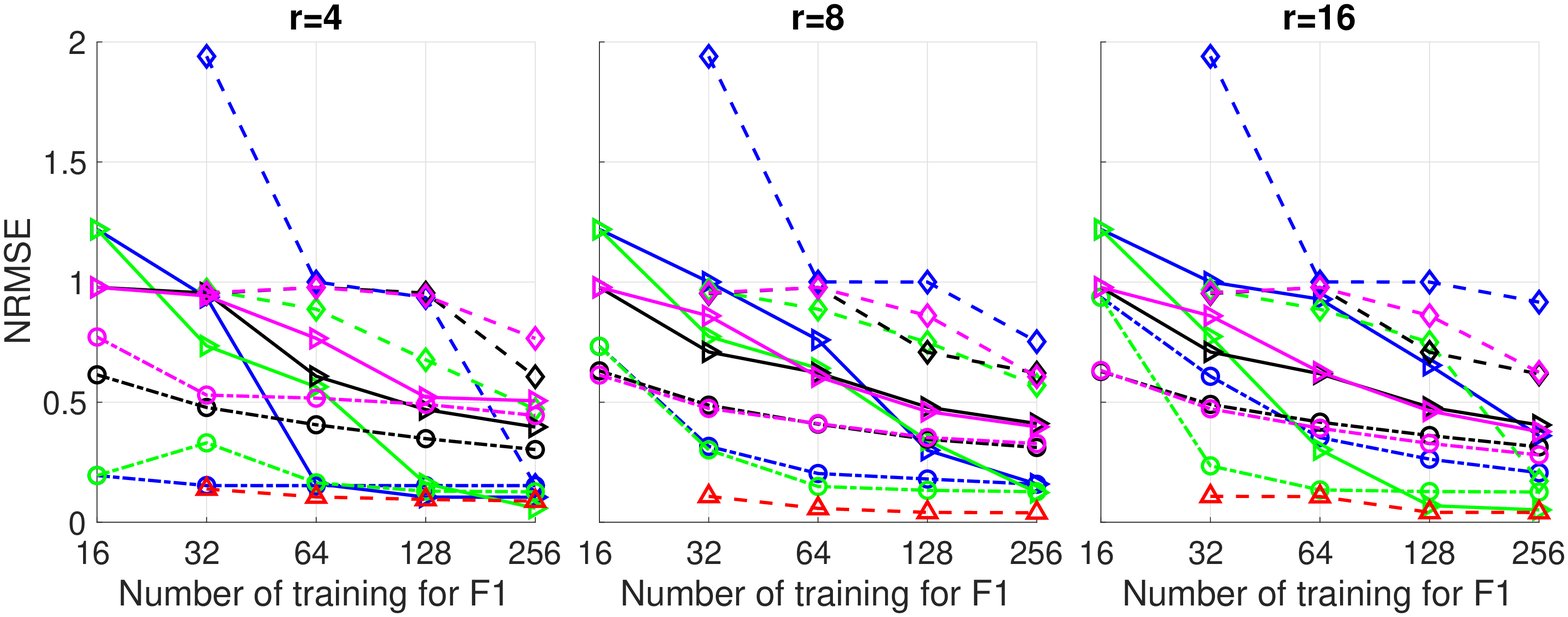}
	\end{subfigure} \\
	\begin{subfigure}[t]{0.9\textwidth}
		\centering
		\includegraphics[width=\textwidth]{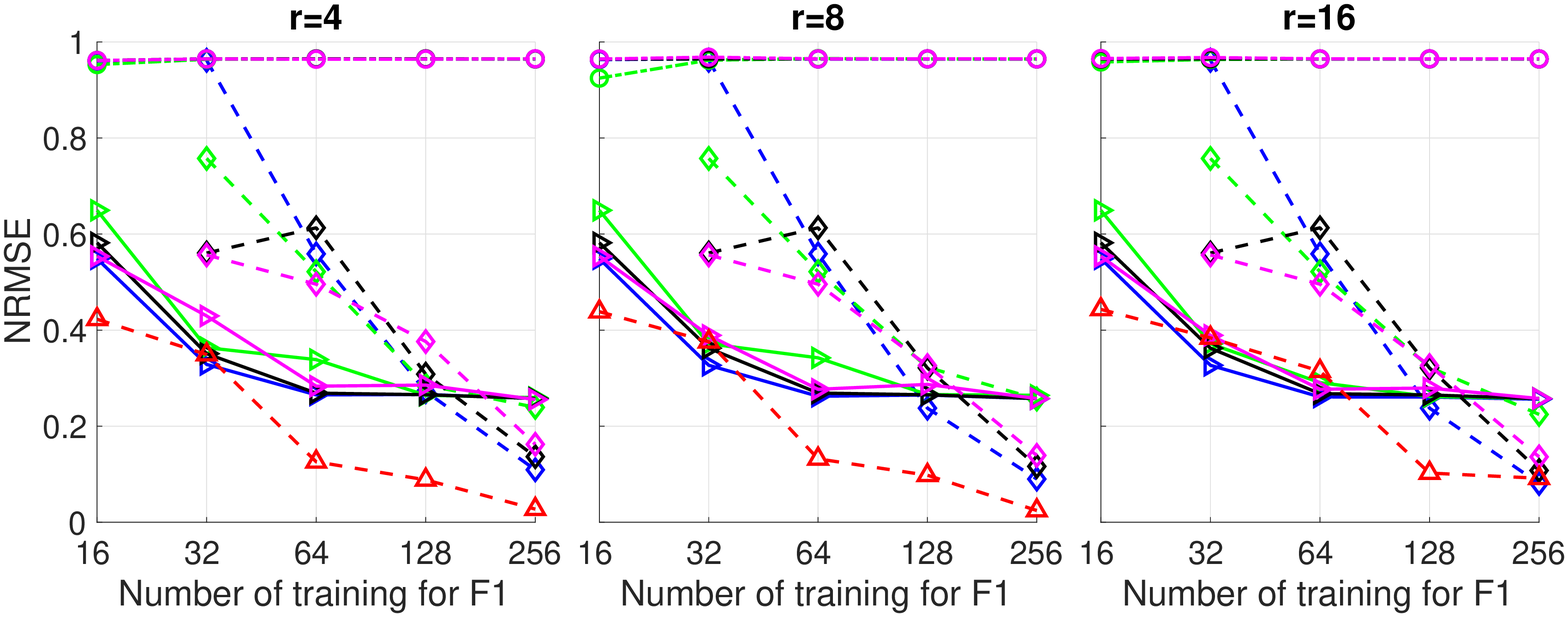}
	\end{subfigure} \\
	\begin{subfigure}[t]{0.9\textwidth}
		\centering
		\includegraphics[width=\textwidth]{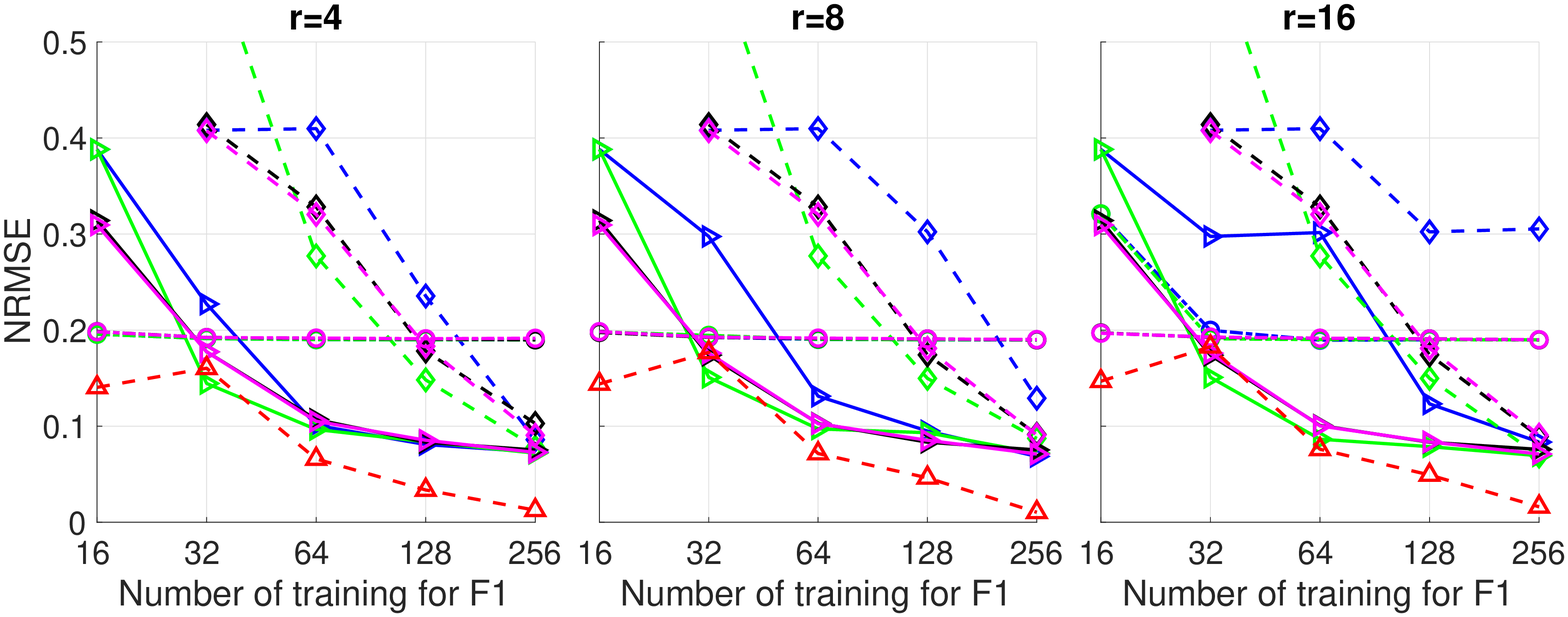}
	\end{subfigure} \\
	\caption{RMSE for Poisson's equation (top row), the heat equation (middle row) and Burger's equation (bottom row) with data of 3 fidelity, low-rank $R=\{4,8,16\}$, and fidelity ratio of $16:4:1$.}
	\label{fig:rmse 3lv_1}
\end{figure}

\begin{figure}[htbp]
	\centering
	\begin{subfigure}[t]{1\textwidth}
		\centering
		\includegraphics[width=\textwidth]{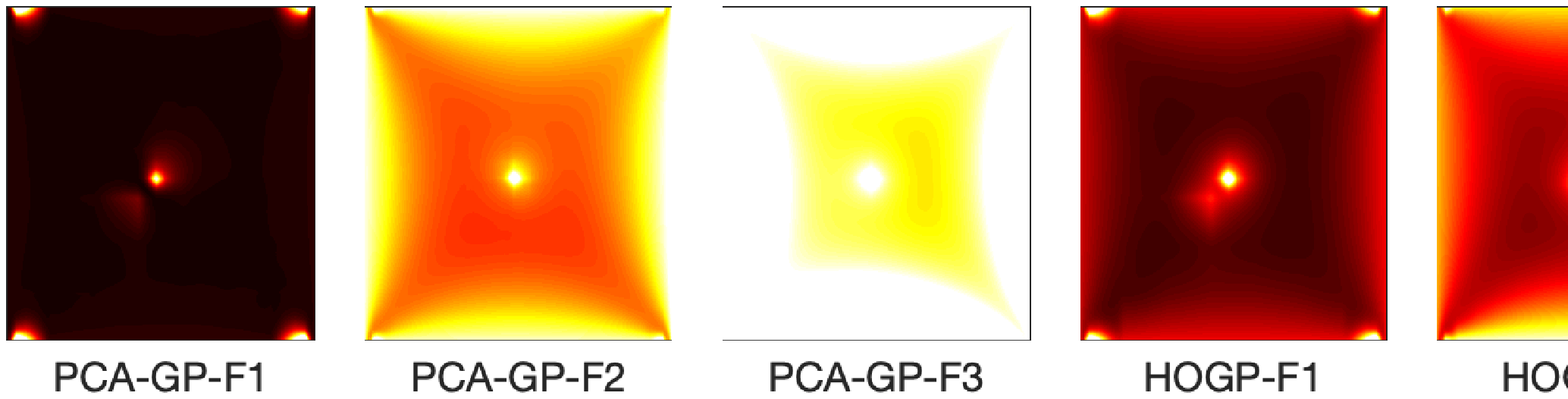}
	\end{subfigure} \\
	\begin{subfigure}[t]{1\textwidth}
		\centering
		\includegraphics[width=\textwidth]{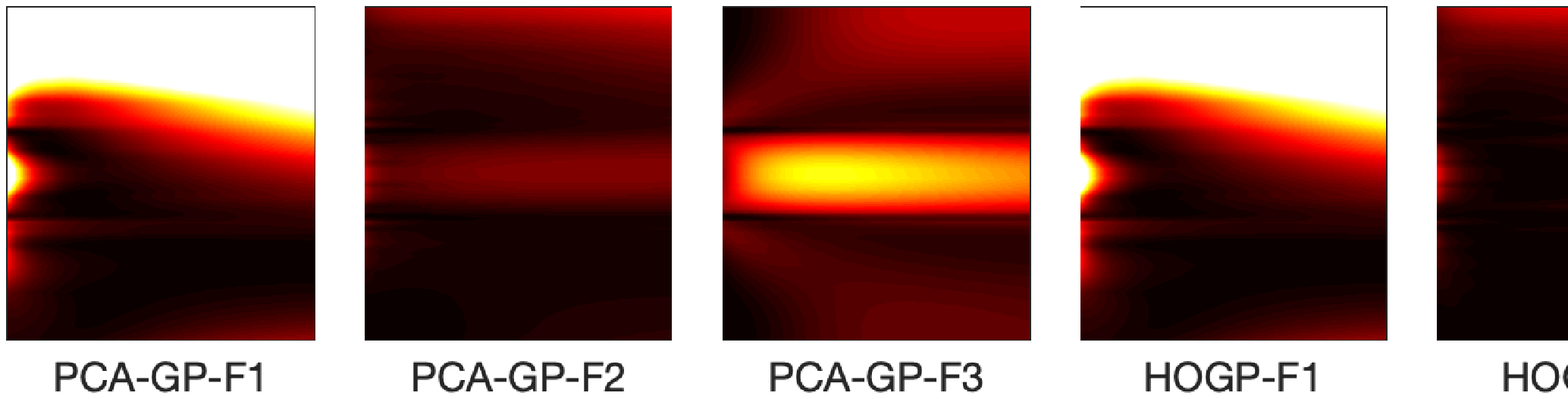}
	\end{subfigure} \\
	\begin{subfigure}[t]{1\textwidth}
		\centering
		\includegraphics[width=\textwidth]{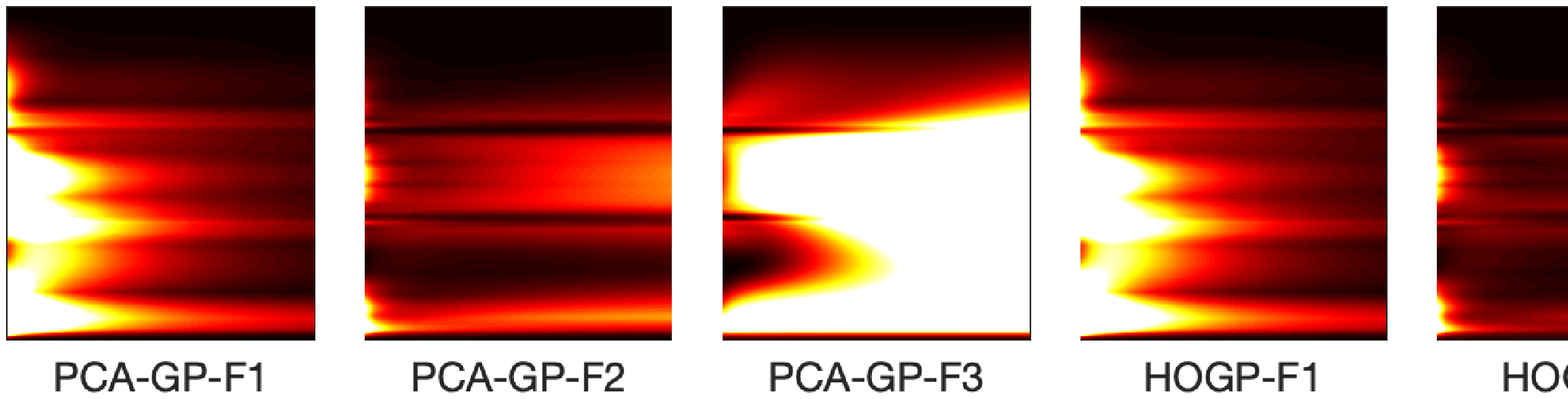}
	\end{subfigure} \\
	\caption{MAE field for Poisson's equation (top row), the heat equation (middle row) and Burger's equation (bottom row) using 64 fidelity-1 samples, 16 fidelity-2 samples, and 4 fidelity-3 samples with row-rank $R=8$.}
	\label{fig:local 3lv_1}
\end{figure}

\subsection{Applications to Real Simulation Problems}
Next, we applied \ours in a more difficult and large-scale physical simulation problem of fluid dynamics. 
Specifically, we emulated the spatial-temporal pressure field generated in a square $2$-d cavity $[0,1]\times [0,1]$ filled with liquid water that is driven by the top boundary representing a sliding lid.
The problem is governed by the incompressible dimensionless Navier-Stokes equations~\citep{chorin1968numerical}:   
\begin{equation} \label{eq:NS}
\frac{\partial \u }{\partial t } (\u \cdot \nabla) \u - Re^{-1} \nabla^2 \u + \nabla {\textit{p}}=0,\vspace{2mm}\quad \nabla \cdot \u=0,
\end{equation}
where $\u=(u_{1},u_{2})^T$ is the liquid velocity, $p$ is the liquid pressure and $Re$ is the Reynolds number. The Navier-Stokes (NS) equation is known computationally challenging and has been a frequent benchmark problem for surrogate models~\citep{perdikaris2017nonlinear,xing2016manifold}.


\noindent \textbf{Simulation setting.} 
The Reynold's number and lid velocity were used as input parameters: $\x=(Re,u_{lid})^T\in [10,1000]\times [0,1]$. All other parameters were kept at the default values.
We used the Sobol sequence to generate $256$ design inputs for training and uniform sampling to generate $64$ inputs for testing.
For each input $\x_i$, we computed five fidelity pressure fields based on five spatial solver grids, which are $8\times8$, $16\times16$, $32\times32$,$64\times64$, and $128\times128$.
We used finite difference on a staggered grid with implicit diffusion, a Chorin projection for the pressure~\citep{seibold2008compact}, and $5000$ fixed time steps to solve the PDEs.
Each pressure field was recorded using a $100\times100\times100$ regular spatial-temporal grid, leading to a 1 million dimension vector.
According to our experiments, the average computational costs for generating a result of different fidelity were $0.3519s, 0.5615s, 1.4449s, 5.7714s$ and $32.2289s$ for the fidelity-1 to fidelity-5 results respectively.
As mentioned previously, in real applications, the low-rank $R$ is normally determined by the preserved variance of the data. 
To reflect this choice, $R$, form here on, denotes the ratio of preserved variance, which implicitly decides the number of bases being used.  
HOGP requires significantly greater computational cost for very high dimensional-output and thus was not included in the following experiments.  

We first conducted the similar two- and three-fidelity experiments as before to examine the model capacities.
The two-fidelity test was done with fixed $256$ fidelity-1 samples; the resulting RMSEs are shown in Fig.~\ref{fig:rmse ns 2lv}.
Results of the three-fidelity test with a fidelity ratio of $16:4:1$ are shown in Fig.~\ref{fig:rmse ns 3lv}. 
It is obvious that 
\ours outperforms others methods in most cases except for the second case in Fig.~\ref{fig:rmse ns 3lv} with 256 fidelity-1 samples.
\begin{figure}[htbp]
	\centering
	\includegraphics[width=\textwidth]{./fig_v2/legendx3.eps}
	\includegraphics[width=\textwidth]{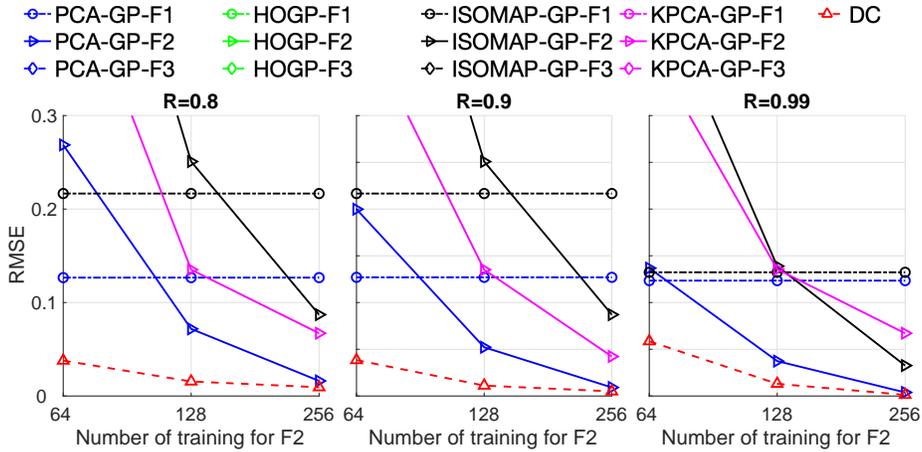}
	\caption{RMSE for lid-driven cavity using training data from two fidelities. The number of fidelity-2 samples varies from $\{64,128,256\}$, whereas the number of fidelity-1 samples is fixed at 256. The low-rank $R$ varies from $\{0.8,0.9,0.99\}$}.
	\label{fig:rmse ns 2lv}
\end{figure} 

\begin{figure}[htbp]
	\centering
	\includegraphics[width=\textwidth]{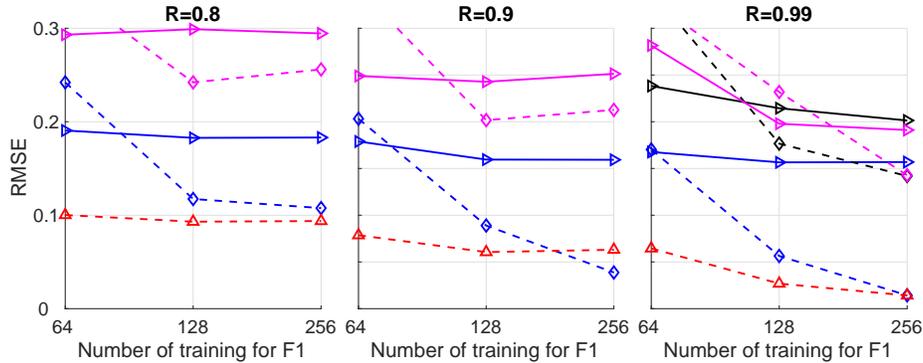}
	\caption{RMSE for lid-driven cavity using three fidelity data. Fidelity ratio is $16:4:1$, and the low-rank $R$ varies from $\{0.8,0.9,0.99\}$.}
	\label{fig:rmse ns 3lv}
\end{figure} 


Finally, we included samples from five fidelities to demonstrate our model performance when dealing with a more realistic application.
Low-fidelity-data-based methods, i.e., the competing methods using fidelity-1 to fidelity-4 data, are not demonstrated since they all show high errors even with 256 training samples.
All methods use a low-rank variance ratio of $R=0.99$.
For \ours, we used two strategies to supply the training samples at each fidelity.
DC(1) uses a fidelity ratio of $16:8:4:2:1$ whereas DC(2) $256:64:16:4:1$.
We increased the training samples at fidelity-5 (and consequently the number of training samples at other fidelities).
The maximum number of training point for each fidelity was restricted to 256 due to the size of our dataset.

Fig.~\ref{fig:rmse ns 5lv} demonstrates the RMSE as a function of simulation cost (in hours), which is the computational cost to generate the training data.
It is clear that \ours can achieve better performance with much less computational cost compared with other methods.
At the simulation cost of about 0.5 hours, \ours can achieve performance that other methods cannot beat even with about five times more simulation cost.
Note that fidelity ratio has only small influence at low simulation cost in terms of model accuracy.
PCA-GP is overall a stable method, but it requires many expensive high fidelity data to slowly improve. 
ISOMAP-GP and KPCA-GP do not perform well even compared with PCA-GP.
According to \citet{xing2015reduced,xing2016manifold}, these two methods have a slow improvement after the low-rank exceeds a certain value, which is clearly the case here as we are preserving $99\%$ of output variance.
Another interesting discovery is that \ours does not just converge quicker but also converges to a much lower error bound.
According to \citet{perdikaris2017nonlinear}, due to the Markov properties, the high-fidelity predictions can benefit from the low-fidelity predictions only with the same input.
When using 256 samples for all fidelities, we should see no improvement compared to directly using an LMC on $256$ fidelity-5 samples, \ie the PCA-GP-F5. 
We believe this is due to the residual additive structure, which not only learns correlations between fidelities but also improves model capacity via the deep structure.  
The MAE as a function of time at the cost of around 0.7 hour is shown in Fig.~\ref{fig:local ns 5lv}.
KPCA-GP-F5 and ISOMAP-GP-F5's error accumulates as time increased and propagates from the top (the driven lid) to the bottom.
PCA-GP-F5 shows much better performance in reducing the error; the error is significant at the beginning first second as due to the drastic changing of the pressure field.
In contrast, \ours show more than an order of magnitude lower error compared to PCA-GP-F5 before $t=4s$, after which the error increases possibly due to the lack of high-fidelity samples to capture the evolving dynamics.

\begin{figure}[h]
	\centering
	\includegraphics[width=0.75\textwidth]{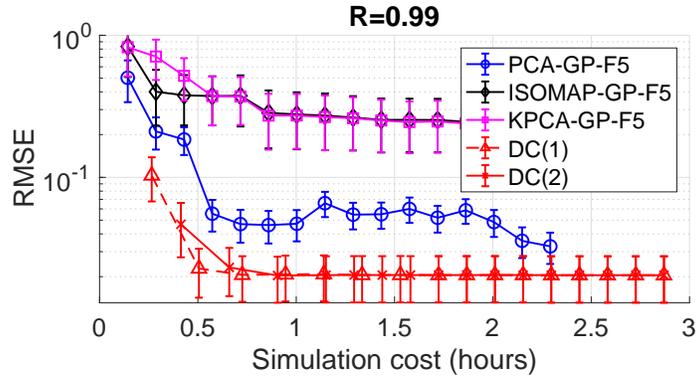}
	\caption{RMSE as a function of computational cost of generating training data.}
	\label{fig:rmse ns 5lv}
\end{figure} 

\begin{figure}[h]
	\centering
	\includegraphics[width=0.45\textwidth]{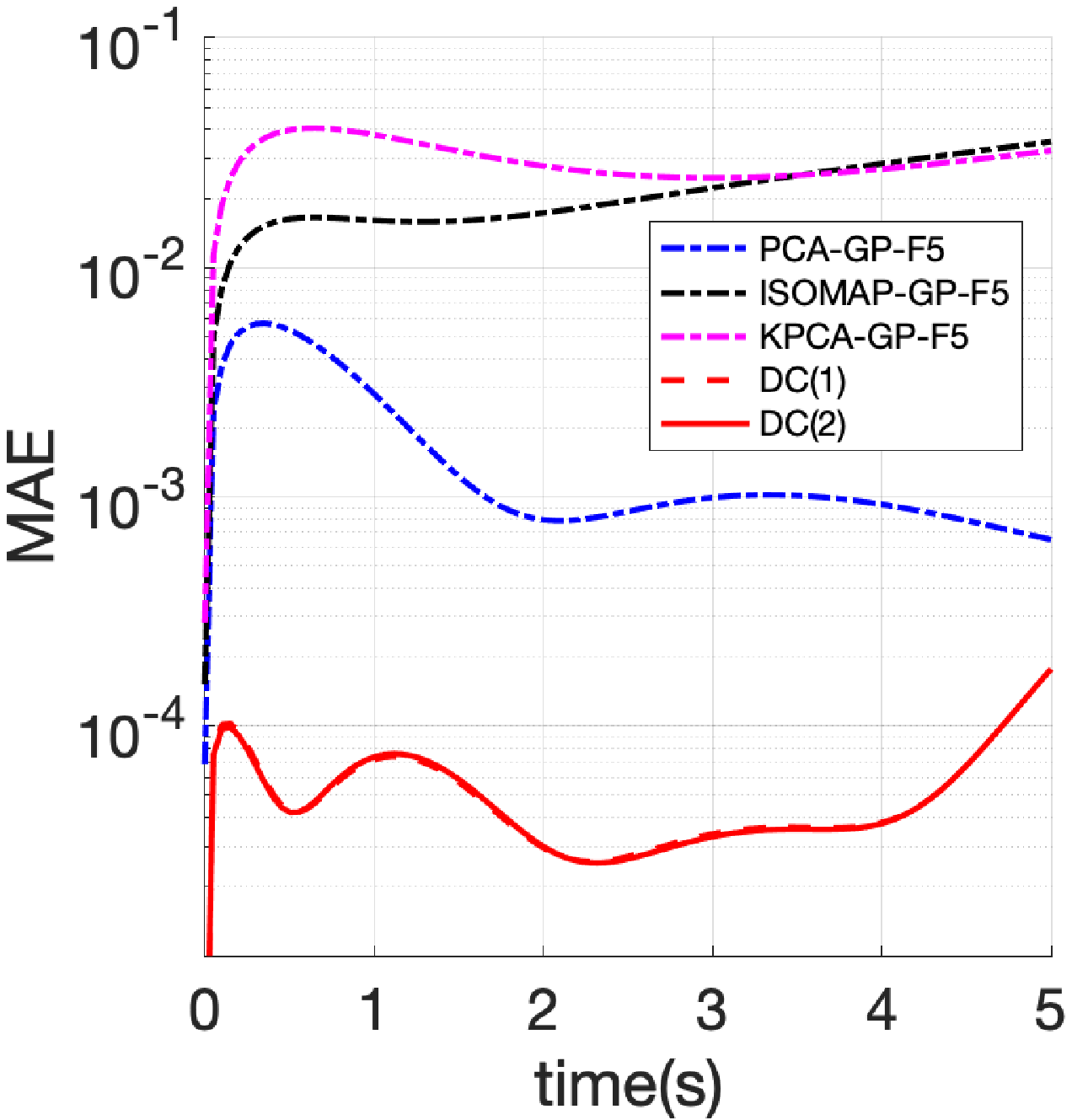}	
	\includegraphics[width=0.4\textwidth]{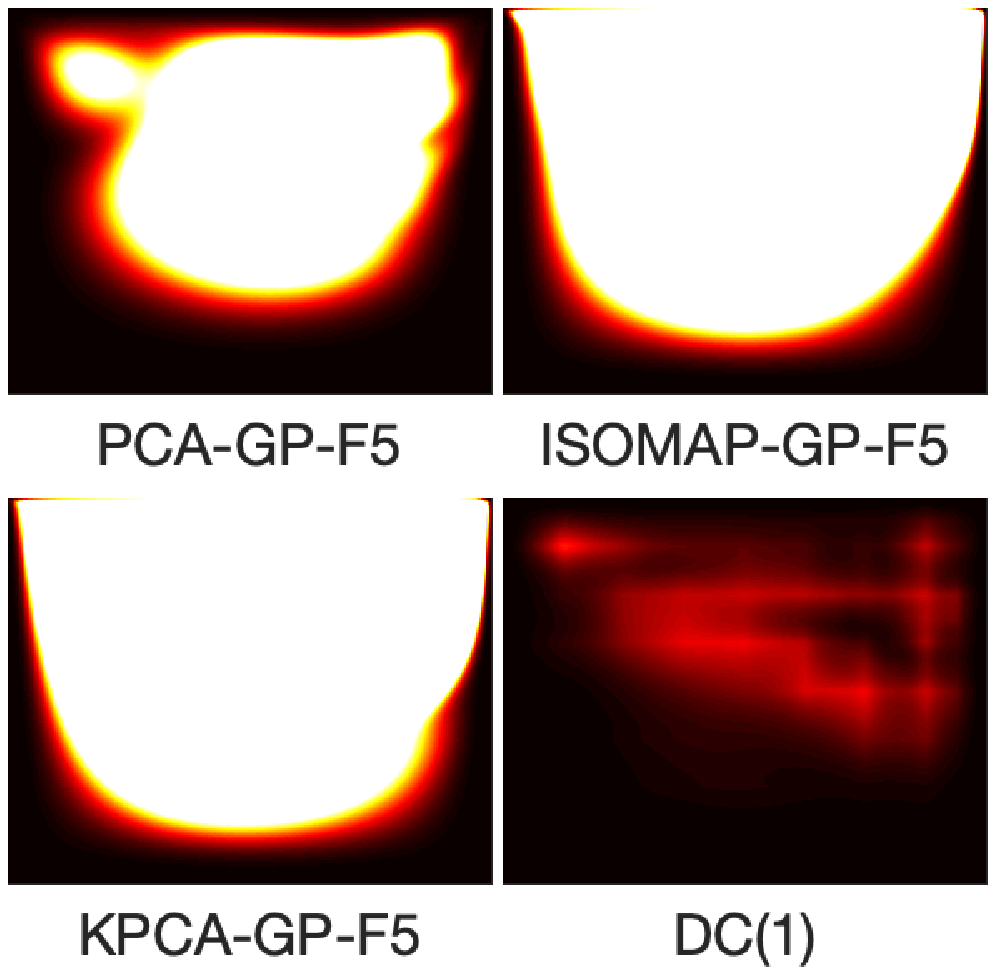}
	\hspace{-0.2in}
	\includegraphics[width=0.1\textwidth]{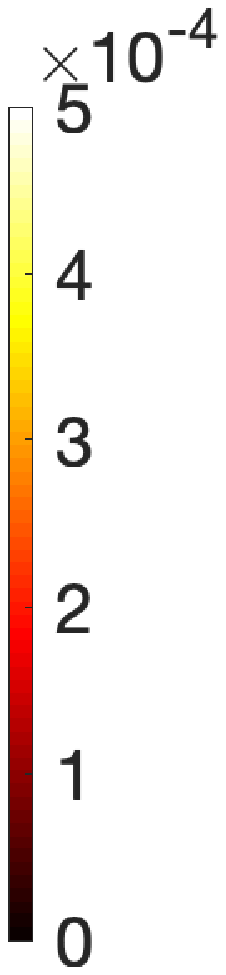}
	\vspace{-0.0in}
	\caption{MAE as a function time (left) and the MAE fields (right) at $t=4$ using a simulation cost of approximately 0.7 hour and low-rank $R=0.99$.}
	\label{fig:local ns 5lv}
\end{figure}


\section{Discussion and Conclusion}
We have presented a novel framework for efficient emulations of spatial-temporal fields of an expensive simulator by utilizing its multi-fidelity setting.
Our model can be seen as a fundamental generalization of the classic autoregressive model for high-dimensional problems based on the general assumption of LMC.
To focus on the model framework itself, our implementation is based on a simple yet successful two-stage method~\citep{higdon2008computer}.
Nevertheless, \ours is also a natural extension of LMC for multi-fidelity data; thus, it is readily implemented with the state-of-the-art methods based on LMC.
For instance, joint learning of the bases can be implemented as suggested in
\citet{goovaerts1997geostatistics}, \citet{bonilla2008multi}, and \citet{alvarez2012kernels}
to potentially improve the model performance.
Tensor decomposition can be applied to the high-dimensional outputs to improve the model scalability when conducting a joint learning~\citep{zhe2019scalable}.
To improve model flexibility, the bases can be assumed functions of the inputs $\x$~\citep{wilson2011gaussian}.
The process convolution~\citep{higdon2002space,boyle2005dependent} can be implemented for non-stationary output correlations.
Manifold learning~\citep{xing2015reduced,xing2016manifold} can be used to derive implicit bases to capture the nonliner output correlations.
Sparse GPs can be implemented for the latent process to reduce computational cost when dealing with large datesets~\citep{snelson2005sparse,hensman2013gaussian}.

The other contributions of this work include the proposed ResPCA algorithm, which presents a new dimension reduction treatment for multi-fidelity data based on an addictive structure across different fidelity.
This method can be potentially used for efficient base extractions for applications such as reduced order models~\citep{efe2003proper,Audouze2012,Hay2010} and sensitivity analysis~\citep{triantafyllidiis2018probabilistic,Saltelli} when multi-fidelity data is available.

The proposed method, \ours, consistently shows superior performance compared to the state-of-the-art GP based high-dimensional emulators.
Considering the extremely high computational cost for high-fidelity simulations in real-life applications, we believe the proposed method should serve as a baseline for emulations of spatial-temporal fields.





\bibliographystyle{unsrt}

\section*{Acknowledgment}
This work has been supported by DARPA TRADES Award HR0011-17-2-0016. Authors would like to thank Kyli McKay-Bishop for proofreading the manuscript.

\section*{References}
\bibliography{./MFSTGP.bib}

\clearpage

\end{document}